\documentclass[a4paper,11pt]{article} 
\usepackage{jcappub}

\usepackage{bm}    
\usepackage{amsmath,amssymb}
\usepackage{color}

\title{Gravitational wave spectrum from kinks on infinite cosmic superstrings with Y-junctions}
\author[a]{Yuka Matsui,}
\author[b]{Koichiro Horiguchi,}
\author[c]{Daisuke Nitta}
\author[a,d]{and Sachiko Kuroyanagi}

\affiliation[a]{Department of physics and astrophysics, Nagoya University, Nagoya, 464-8602, Japan}
\affiliation[b]{Medical system division, Amelieff, Tokyo, 108-0014 Japan}
\affiliation[c]{Sendai, Miyagi, 980-0044, Japan}
\affiliation[d]{Instituto de F\'isica Te\'orica, Universidad Auton\'oma de Madrid, 28049 Madrid, Spain}

\emailAdd{matsui.yuka@f.mbox.nagoya-u.ac.jp}
\emailAdd{horiguchik@amelieff.jp}
\emailAdd{nittadaisuke2112@gmail.com}
\emailAdd{skuro@nagoya-u.jp}

\abstract{
  We calculate the gravitational wave (GW) background spectra from kink propagation and kink-kink collisions on infinite cosmic superstrings.
We take into account two characteristics of the cosmic superstring network: a small reconnection probability and Y-junctions. First, a small reconnection probability increases the number of infinite strings inside the horizon and enhances the kink production, which leads a larger amplitude of the GW background. Second, a kink going through a Y-junction transforms into three daughter kinks. In this way, the existence of Y-junctions also increases the number of kinks on cosmic superstrings. However, at the same time, it smooths out the sharpness of kinks rapidly and reduces the number of sharp kinks, which are responsible for the emissions of strong GW bursts. We compute the number distribution of kinks as a function of the sharpness by taking into account the above two effects, and translate it to the amplitude of the GW background spectra. We first investigate the case of the string network with equal string tensions, and find that the effect of Y-junctions to smooth out kink sharpness dominates that of the enhancement of the kink number by a small reconnection probability, and the GW amplitude turns out to be smaller than the ordinary cosmic string case. On the other hand, for non-equal string tensions, we find that there is a parameter space where the GW amplitude is slightly enhanced by the effect of a small reconnection probability. 
}

\begin{document}
\maketitle

\section{Introduction}
Gravitational waves (GW) are a new and essential tool in the study of astrophysical and cosmological phenomena.
One of the interesting sources from the early Universe is one dimensional topological defects formed by the spontaneous symmetry breaking, called cosmic strings \cite{Kibble:1976sj, Vilenkin}. 
Superstring theory predicts that fundamental strings (F-strings) and 1-dimensional Dirichlet branes (D-strings) might be stretched to cosmic size and play the role of cosmic strings \cite{Sarangi:2002yt, Jones:2003da, Dvali:1998pa, Dvali:2003zj}.
This gives rise to the possibility that cosmic string observations could open new ways to access high energy physics such as string theory and brane inflation models.

Cosmic strings are known to reconnect when they intersect. 
Loops are formed when infinite strings, whose length is larger than the Hubble horizon, reconnect. They shrink by transferring their energy to GWs and eventually disappear. 
While the number and the total length of infinite strings inside the horizon increase as the horizon extends, infinite strings can reduce their length inside the Hubble horizon by forming loops. Because of the balance between the two effects, the number of infinite strings inside the horizon is always kept constant, which is known as the scaling law, and the energy density of strings does not interrupt the standard evolution of the Universe.

The field-theoretic cosmic strings and cosmological size strings motivated by superstring theory (hereafter we call them ``cosmic strings'' and ``cosmic superstrings'') have slightly different features. First, cosmic string networks can be made of different string types, D-strings and F-strings, and they have different tensions. Their reconnection probability $p$ could be much smaller than unity, while $p\sim 1$ is expected for cosmic strings \cite{Shellard:1987bv}. The value of $p$ can range as $10^{-1} \lesssim p_{\rm D} \lesssim 1$ for D-strings and $10^{-3} \lesssim p_{\rm F} \lesssim 1$ for F-strings, respectively \cite{Jackson:2004zg}. Furthermore, D-strings and F-strings can form bound states and the string network can contain Y-junctions where two different strings join and separate. 

Sharp structures, called kinks, are produced at the point where two strings reconnect. They emit GWs through their propagation \cite{Damour:2000wa, Damour:2001bk} as well as through their collisions \cite{Binetruy:2009vt}. Singularity points typically arise on loops, called cusps, that also emit strong GW bursts. These GWs overlap one another and form a GW background. 
The GW background from cusps on cosmic string loops has been increasingly studied in the literature
\cite{Damour:2004kw, Siemens:2006yp, DePies:2007bm, Olmez:2010bi, Sanidas:2012ee, Sanidas:2012tf, Binetruy:2012ze, Kuroyanagi:2012wm, Kuroyanagi:2012jf, Siemens:2002dj, Polchinski:2006ee}.
For cosmic superstrings, \cite{Sousa:2016ggw} studied the GW background from cusps on loops by taking into account the different tensions and reconnection probabilities, while \cite{Binetruy:2010cc} investigated GWs from kinks on loops with the effect of Y-junctions. Interestingly, it was pointed out that the number of kinks increases when they enter Y-junctions \cite{Binetruy:2010bq} and could increase the amplitude of the GW background \cite{Binetruy:2010cc}. While loops can generate GWs of wavelength shorter than the loop size, infinite strings generate GWs over a wide range of wavelengths from the Cosmic Microwave Background (CMB) to the interferometric scales, although the amplitude had been considered to be much smaller than the one from loops \cite{Sakellariadou:1990ne,Hindmarsh:1990xi,Figueroa:2012kw,Kawasaki:2010yi,Matsui:2016xnp}. However, in our previous work, we estimated the amplitude of GW background from kink propagation and kink-kink collisions on infinite strings \cite{Matsui:2016xnp,Matsui:2019obe}, and found that kink-kink collisions on infinite strings generate a GW background with a relatively large amplitude, which is almost comparable to the one from loops.

In this paper, we further investigate the GW background from kink propagation and kink-kink collisions on infinite strings by taking into account the characteristics of cosmic superstrings such as a small reconnection probability and Y-junctions. These effects first appear in the calculation of the correlation length, which gives the number of infinite strings inside the horizon. We adopt the extended velocity-dependent evolution equations developed in \cite{Avgoustidis:2007aa,Avgoustidis:2009ke}, which include the effects of the formation of junctions between strings of different tensions. We also take into account the fact that a small reconnection probability reduces the efficiency of loop formation. Secondly, based on \cite{Copeland:2009dk}, we extend the evolution equation for the number distribution of kink sharpness to include the effect of the kink proliferation due to the fact that a kink passing through a Y-junction transforms into three kinks: a reflected kink and two transmitted kinks \cite{Binetruy:2010bq}. By numerically solving the evolution equations, we obtain the kink number density and translate it into the amplitude of the GW background. We also consider the effect of the backreaction of a large GW emission, which was formulated in \cite{Matsui:2019obe}, in order to avoid overestimation of the GW amplitude. 

The structure of this paper is as follows. In section \ref{sec:dynamics} and \ref{sec:network}, we review the network dynamics of cosmic strings and superstrings with Y-junctions. Then, in section \ref{sec:kink_distribution}, we derive the time evolution equation for the distribution function of kinks on infinite cosmic superstrings, and solve it numerically.
In section \ref{sec:propagating_kink}, we calculate the GW background spectrum from propagating kinks on infinite cosmic superstrings with Y-junctions.
In section \ref{sec:kink-kink_collision}, we calculate the GW background spectrum from kink-kink collisions when considering the gravitational emission effect.
Section \ref{sec:summary} is devoted to a summary.

\section{Dynamics of strings} \label{sec:dynamics}
Let us consider dynamics of a cosmic string and a cosmic superstring in Friedmann-Lema\^itre-Robertson-Walker (FLRW) metric,
\begin{equation}
  {\rm d}s^2 = -{\rm d}t^2 +a(t)^2 {\rm d}{\bm x}^2 = g_{\mu \nu} {\rm d}x^\mu {\rm d}x^\nu. 
  \label{eq:FLRW_metric}
\end{equation}
The track of a string makes a two-dimensional world sheet and the string position is described as 
\begin{equation}
  x^\mu = x^\mu(\sigma^a),
\end{equation}
where $a = 0, \, 1$ and $\sigma^0$ and $\sigma^1$ are timelike and spacelike world sheet coordinates, respectively.
The Nambu-Goto action is a good approximation to describe the behavior of the string \cite{Vilenkin:1984ib},
\begin{equation}
  S[x^\mu] = -\mu \int {\rm d}^2 \zeta \sqrt{-{\rm det} (\gamma_{ab})} , 
  \label{eq:string_action}
\end{equation}
where $\mu$ is the tension of string, and $\gamma_{ab} = \frac{\partial x^\mu}{\partial \sigma^a} \frac{\partial x^\nu}{\partial \sigma^b} g_{\mu \nu}$ is the metric on the world sheet, called induced metric.
Varying the action Eq.~\eqref{eq:string_action} with respect to $x^\mu$, we obtain the equation of motion of the string 
\begin{eqnarray}
  \ddot{{\bm x}} + 2\frac{\dot{a}}{a} & \dot{{\bm x}} & (1 -\dot{{\bm x}}^2) = \frac{1}{\epsilon} \left (\frac{{\bm x}'}{\epsilon} \right )',  \label{eq:string_EoM_FLRW} \\
  \epsilon & \equiv & \sqrt{\frac{{{\bm x}'}^2}{1-\dot{{\bm x}}^2}}, 
  \label{eq:epsilon}
\end{eqnarray}
where the dot denotes the derivative with respect to the 
conformal time $\sigma^0$, the prime denotes the derivative with respect to $\sigma^1$, and $a$ is the scale factor of the Universe.
The variable 
\begin{equation}
   {\bm p}_{\pm} \equiv \dot{{\bm x}} \mp \frac{1}{\epsilon} {\bm x}', 
   \label{eq:p_pm}
\end{equation}
corresponds to the left and right moving modes on strings.

When cosmic strings intersect one another, they reconnect and produce sharp structures, called kinks.
In the case of cosmic superstrings, strings reconnect with the probability of $p\ll 1$ and kinks are produced if they reconnect, otherwise they pass through because of the extra dimension.
Kinks move with the speed of light and its sharpness decreases as the Universe expands.
The sharpness of a kink is defined by 
\begin{equation}
  \psi \equiv \frac{1}{2} (1- {\bm p}_{+, \, 1} \cdot {\bm p}_{+, \, 2}),  
  \label{eq:psi_define}
\end{equation}
where the number of the suffix denotes the modes on the right and left of the kink and ${\bm p}_{\pm}$ relates to the mean square velocity of strings $v$ as $\left < {\bm p}_{+} \cdot {\bm p}_{-} \right > = -(1-2v^2) \equiv -\kappa$.
One can rewrite Eq.~\eqref{eq:string_EoM_FLRW} in terms of $\psi$ using Eq.~\eqref{eq:psi_define}, and get the evolution equation of the sharpness
\begin{equation}
\dot{\psi} = -2H \kappa \psi,
\end{equation}
where $H = ({\rm d}a/{\rm}dt)/a$ is the Hubble parameter. Then we find the solution 
 \begin{equation}
   \psi \propto t^{-2 \zeta},
   \label{eq:zeta}
 \end{equation}
 where $\zeta \equiv (1-2v^2) \nu$ with $\nu$ describing the time evolution of the scale factor as $a \propto t^{\nu}$. The parameter $\zeta$ characterizes the speed of the sharpness decreases by the expansion of the Universe \cite{Copeland:2009dk}.
   We numerically obtain the value of $\nu$ using $\nu=\frac{\ln{(a/a_i)}}{\ln{(t/t_i)}}$ where $a_i$ and $t_i$ are the initial scale factor and the initial time.

 \section{Evolution of the string network} \label{sec:network}
 \subsection{The case of ordinary cosmic strings}\label{sec:network_cosmicstring}
 Cosmic string network is known to follow the scaling law, where the correlation length grows in proportion to the horizon size of the Universe $\sim t$. In other words, the number of infinite strings inside the horizon is conserved in time.
In the velocity-dependent one-scale model (VOS model) \cite{Martins:1996jp}, the evolution of the infinite string network is characterized by the correlation length $L$ and the average velocity of strings $v$. The evolution of $L$ and $v$ are obtained by simultaneously solving the following equations \cite{Kibble:1984hp}, 
\begin{eqnarray}
  \frac{{\rm d} L}{{\rm d} t} & = & HL(1+v^2) +\frac{1}{2}c v,  \label{eq:L_eq} \\
  \frac{{\rm d} v}{{\rm d} t} & = & (1-v^2) \left (\frac{k(v)}{L} -2Hv  \right ),  \label{eq:v_eq}
\end{eqnarray}
where $k(v) = \frac{2 \sqrt{2}}{\pi} \frac{1-8v^6}{1+8v^6}$ is the effective curvature \cite{Pogosian:1999np}.
The last term in Eq.~\eqref{eq:L_eq} describes the energy loss into loops and we set the loop chopping efficiency parameter as $c \simeq 0.23$ \cite{Martins:2000cs}.

\begin{figure}[htbp]
  \centering
  \begin{tabular}{c}
    \begin{minipage}{0.5\hsize}
      \centering
      \includegraphics[width=7.5cm,clip]{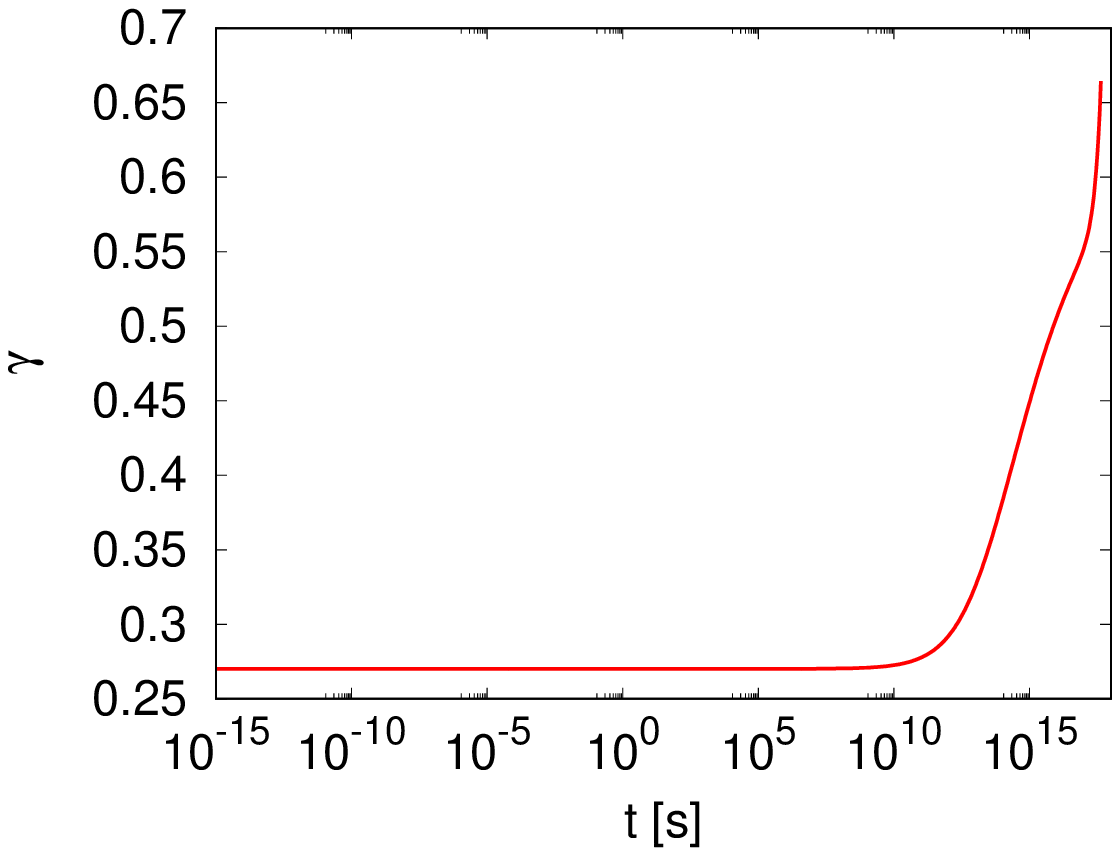}
    \end{minipage}
    \begin{minipage}{0.5\hsize}
      \centering
      \includegraphics[width=7.5cm,clip]{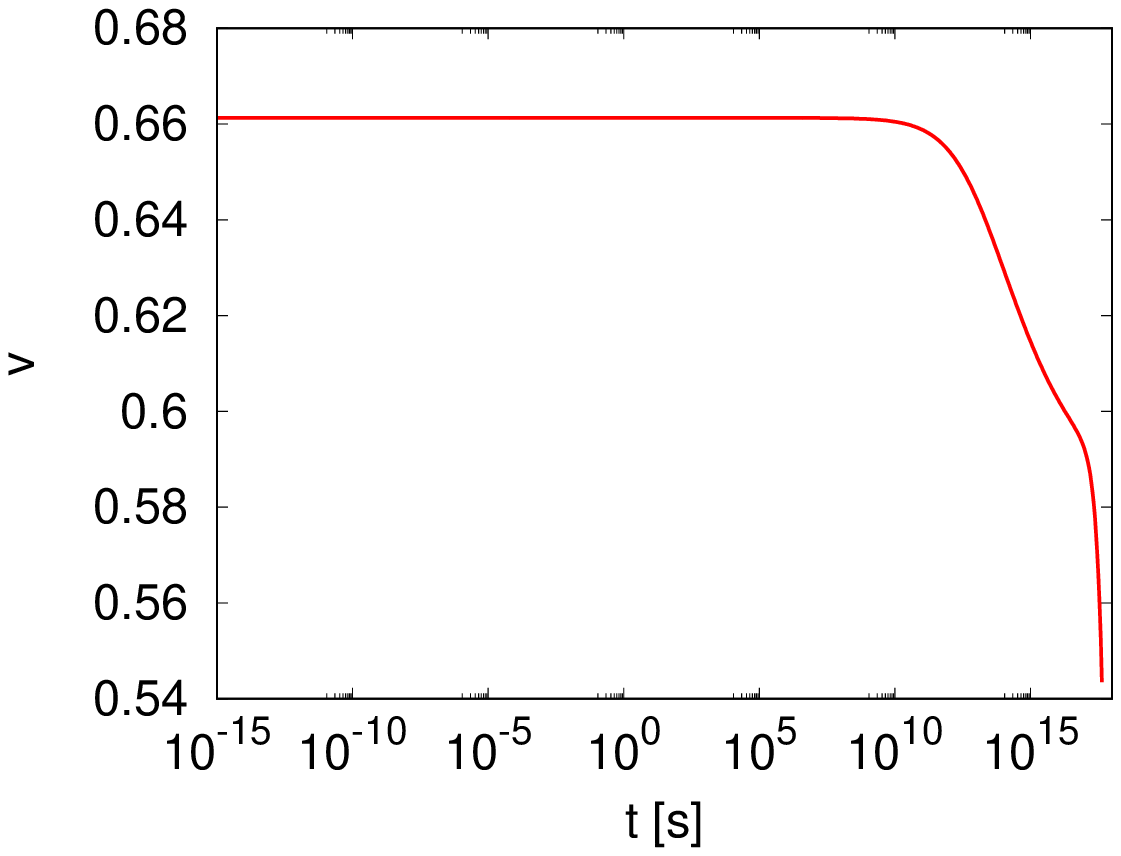}
    \end{minipage}
  \end{tabular}
  \caption{The time evolution of $\gamma$ and $v$, obtained using the VOS model. The horizontal axis is the cosmic time $t$ and the vertical axis is the normalized correlation length $\gamma \equiv L/t$ on the left panel and the average velocity $v$ on the right panel.}	
  \label{fig:scaling_CS}	
\end{figure}

Figure \ref{fig:scaling_CS} shows the time evolution of the correlation length normalized by the horizon size $\gamma \equiv L/t$ and the average velocity $v$, which are obtained by numerically solving Eqs.\eqref{eq:L_eq} and \eqref{eq:v_eq}.
The calculation starts from the radiation-dominated (RD) era, followed by the matter-dominated (MD) and the cosmological constant dominated eras. We find a clear scaling behavior $\gamma\propto {\rm const.}$ in the RD era, while we do not see it in the following eras, since it takes time to reach the scaling regime.
The Hubble parameter is calculated using $H=H_0\sqrt{\Omega_{r}a^{-4}+\Omega_{m}a^{-3}+\Omega_{\Lambda}}$ with the Hubble constant $H_0=100 h$km/s/Mpc.  We use $\Omega_{r}h^2=4.31\times 10^{-5}$ and the cosmological parameters obtained from Planck satellite: $h = 0.678$, $\Omega_{\rm m} = 0.308$ and $\Omega_{\Lambda}= 0.692$ \cite{Ade:2015xua}.

\subsection{The case of cosmic superstrings}
The characteristics of cosmic superstrings affect network evolution. 
First, a smaller reconnection probability reduces the efficiency of loop production and increases the number density of infinite strings. 
Second, the formation process of Y-junctions also affects the scaling behavior and changes the number density of strings. 
Let us consider the situation that strings of types 1 and 2 make bound states and transform into a string of type 3.
These three strings can have different tensions and we label them as $\mu_1, \, \mu_2, \, \mu_3$.
The correlation lengths and the average velocities can be also different for each string type and are labeled as $L_1, \, L_2, \, L_3$ and $v_1, \, v_2, \, v_3$, respectively.
In order to obtain their values, we adopt the extended VOS model developed in \cite{Avgoustidis:2007aa,Avgoustidis:2009ke} (see also \cite{Pourtsidou:2010gu,Sousa:2016ggw} for applications) and
the evolution equations of correlation lengths and velocities are given by
\begin{eqnarray}
  \frac{{\rm d} L_1}{{\rm d} t} & = & HL_1(1+{v_1}^2) +\frac{1}{2}c_1p_1^{n_p}v_1 +\frac{1}{2} \frac{\tilde{d}_{12}^{3} v_{12} {L_1}^2}{L_2(L_1 +L_2)}, \label{eq:L_1_CSS_eq} \\
    \frac{{\rm d} L_2}{{\rm d} t} & = & HL_2(1+{v_2}^2) +\frac{1}{2}c_2p_2^{n_p}v_2 +\frac{1}{2} \frac{\tilde{d}_{12}^{3} v_{12} {L_2}^2}{L_1(L_1 +L_2)},  \label{eq:L_2_CSS_eq} \\
      \frac{{\rm d} L_3}{{\rm d} t} & = & HL_3(1+{v_3}^2) +\frac{1}{2}c_3p_3^{n_p}v_3 -\frac{1}{2} \frac{\tilde{d}_{12}^{3} v_{12} {L_3}^3}{L_1 L_2(L_1 +L_2)},  \label{eq:L_3_CSS_eq} \\
  \frac{{\rm d} v_1}{{\rm d} t} & = & (1-{v_1}^2) \left (\frac{k(v_1)}{L_1}-2Hv_1 \right ), \label{eq:v_1_CSS_eq} \\
  \frac{{\rm d} v_2}{{\rm d} t} & = & (1-{v_2}^2) \left (\frac{k(v_2)}{L_2}-2Hv_2 \right ),  \label{eq:v_2_CSS_eq} \\
  \frac{{\rm d} v_3}{{\rm d} t} & = & (1-{v_3}^2) \Biggl\{\frac{k(v_3)}{L_3}-2Hv_3 +\tilde{d}_{12}^{3} \frac{v_{12}}{v_3} \frac{\mu_1 +\mu_2 -\mu_3}{\mu_3} \frac{L_3^2}{L_1 L_2 (L_1 +L_2)}
     \Biggr\},  \label{eq:v_3_CSS_eq}
\end{eqnarray}
where $v_{12} \equiv \sqrt{v_1^2 +v_2^2}$.
The first terms of the correlation lengths are the same as the cosmic string case. In the second terms, which describe the effect of loop production, we have an additional factor of $p^{n_p}$.
For a simple one-scale model \cite{Martins:1996jp}, we expect $n_p=1$.  However, numerical simulations indicate that small-scale structures on strings increase the loop production efficiency and we can effectively include it by setting $n_p=1/3$ \cite{Avgoustidis:2005nv}, while another simulation indicates $n_p=1/2$ \cite{Sakellariadou:2004wq}.
In this paper, we set $c_1 = c_2 = c_3 \simeq 0.23$ and investigate the two cases: $n_p=1$ and $1/3$.

The parameter $\tilde{d}_{12}^{3}$ describes the efficiency of the process that strings of type $1$ and $2$ produce type $3$, and is given by \cite{Jackson:2004zg,Jackson:2007hn}
\begin{equation}
  \tilde{d}_{12}^{3} = \tilde{d}_{(p,q),(p',q')} P^{\pm}_{(p,q),(p',q')},
\end{equation}
where the probability of Y-junction formation by a $(p, q)$ string and a $(p', q')$ string is given by 
\begin{equation}
  P^{\pm}_{(p,q),(p',q')} = \frac{1}{2} \left (1 \mp \frac{pp'g_s^2 +qq'}{\sqrt{p^2 g_s^2 +q^2} \sqrt{{p'}^2g_s^2 +{q'}^2}}
   \right ),
   \label{eq:P_pm}
\end{equation}
and $g_s$ is the string coupling.
Typically, $\tilde{d}_{(p,q),(p',q')}$ ranges as $10^{-3} \leq \tilde{d}_{(p,q),(p',q')} \leq 1$.
Following \cite{Avgoustidis:2009ke}, in order to implement the kinematic constraint, we rewrite $\tilde{d}_{12}^{3}$ using the suppression factor $S_{12}^{3}$ as 
\begin{equation}
  \tilde{d}_{12}^{3} \to \tilde{d}_{12}^{3} = S_{12}^{3} \tilde{d}_{(p,q),(p',q')} P^{\pm}_{(p,q),(p',q')}.
\end{equation}
Using the string velocity $v$ and the angle of strings $\beta$ at the collision, $S_{12}^{3}$ is written by
\begin{equation}
  S_{12}^{3} = \frac{2}{\pi} \int_{0}^{1}{\rm d}v \int _{0}^{\frac{\pi}{2}} {\rm d} \beta ~ \Theta(-f(v, \, \beta)) {\rm exp} \left(\frac{-(v-v_{12})^2}{\sigma_v^2} \right ) < 1
\end{equation}
where $\Theta$ is the Heaviside function imposing the kinematic constraints
\begin{equation}
  f(v, \, \beta) \equiv A_1(1 -v^2)^2 +A_2(1-v^2) +A_3 < 0.
\end{equation}
The coefficients are given by 
\begin{eqnarray}
A_1 & = & \bar{\mu}_+^2 {\rm cos}^2 \beta (\bar{\mu}_3^2 -\bar{\mu}_+^2 {\rm sin}^2 \beta -\bar{\mu}_-^2 {\rm cos}^2 \beta), \\
A_2 & = & 2 \bar{\mu}_+^2 \bar{\mu}_-^2 {\rm cos}^2 \beta -\bar{\mu}_3^4 -(2{\rm cos}^2 \beta -1) \bar{\mu}_+^2 \bar{\mu}_3^2, \\
A_3 & = & \bar{\mu}_3^4 -\bar{\mu}_+^2 \bar{\mu}_-^2,
\end{eqnarray}
with $\bar{\mu}_\pm = \mu_1 \pm \mu_2$. 
The string of type 3 is bound states of p F-strings and q D-strings and its tension $\bar{\mu}_3$ is given by
\begin{equation}
  \bar{\mu}_3 \equiv \frac{\mu_{\rm F}}{g_s} \sqrt{p^2g_s^2+q^2},
\end{equation}
where $\mu_{\rm F}$ is the tension of the lightest F-string. 
The tension of the type 3 string $\bar{\mu}_3 = \mu_3$ would be roughly determined by heavier strings among type 1 and 2.
For the velocity variance, we take $\sigma_v^2 = 0.25$.

A reconnection probability also changes the probability of string collisions and affects the efficiency parameter $\tilde{d}_{12}^{3}$.  We assume that the efficiency parameter has the following dependency
\begin{equation}
  \tilde{d}_{12}^{3} \to \tilde{d}_{12}^{3} = S_{12}^{3} \tilde{d}_{(p,q),(p',q')} P^{\pm}_{(p,q),(p',q')} p_3^{n_p}.
\end{equation}
In this paper, we consider the simple case where one D-string $(p,q) = (0,1)$ and one F-string $(p',q') = (1,0)$ make the bound state and form Y-junctions.
Substituting $(p,q),(p',q') = (0,1),(1,0)$ into Eq.\eqref{eq:P_pm}, we get $P^{\pm}_{(p,q),(p',q')} = \frac{1}{2}$.
Since we are interested in seeing the maximum effect of Y-junctions, we set
$\tilde{d}_{(p,q),(p',q')} = 1$. 

\begin{figure}[htbp]
  \centering
  \begin{tabular}{c}
  Case A: $\mu_1:\mu_2:\mu_3=1:1:1$, $n_p=1$, $G\mu_1=G\mu_2=G\mu_3=10^{-11}$ \\
    \begin{minipage}{0.5\hsize}
      \centering
      \includegraphics[width=7.5cm,clip]{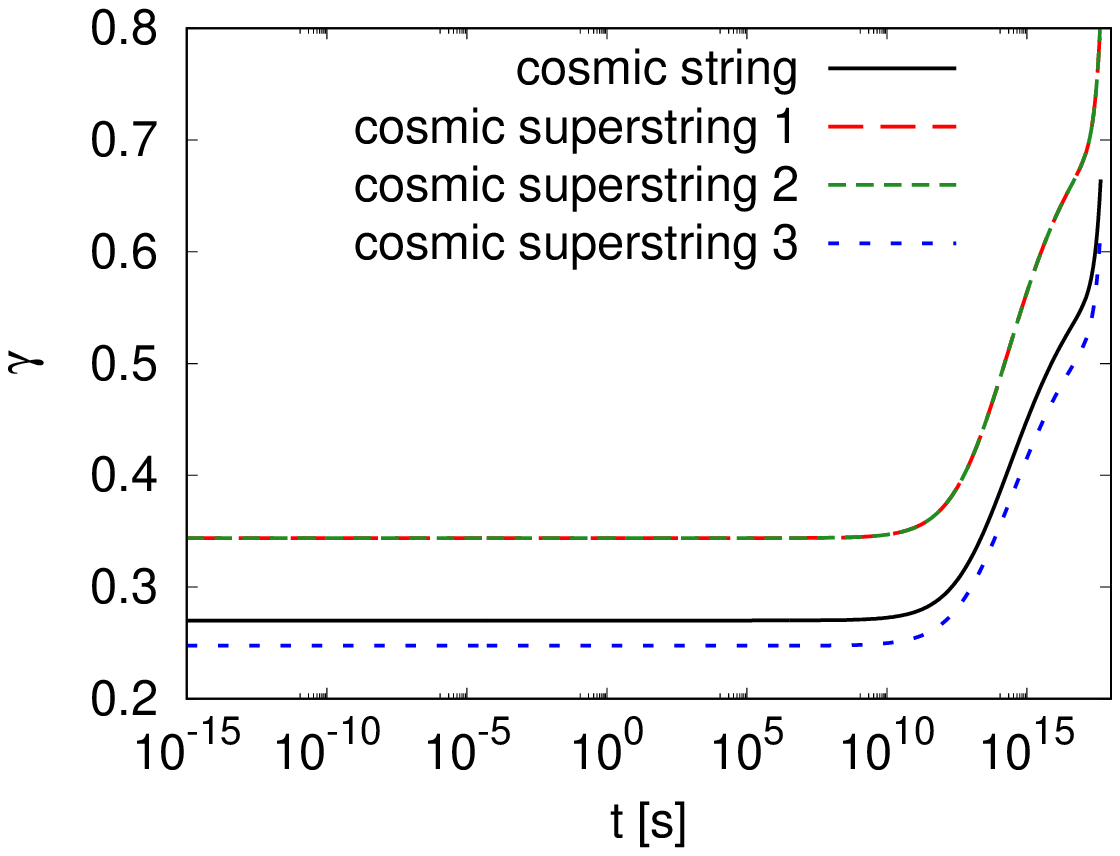}
    \end{minipage}
    \begin{minipage}{0.5\hsize}
      \centering
      \includegraphics[width=7.5cm,clip]{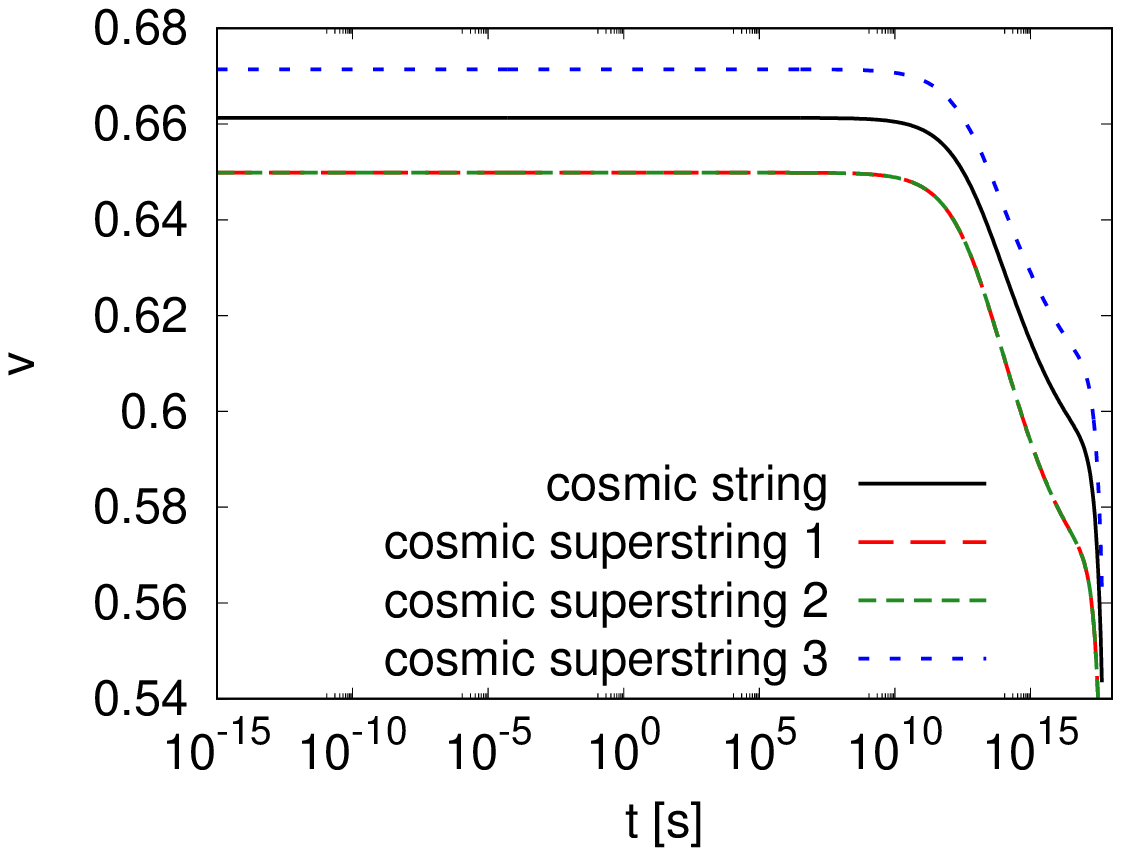}
    \end{minipage} \vspace{20pt} \\
    Case B: $\mu_1:\mu_2:\mu_3=1:1:1$, $n_p=\frac{1}{3}$, $G\mu_1=G\mu_2=G\mu_3=10^{-11}$ \\
        \begin{minipage}{0.5\hsize}
      \centering
      \includegraphics[width=7.5cm,clip]{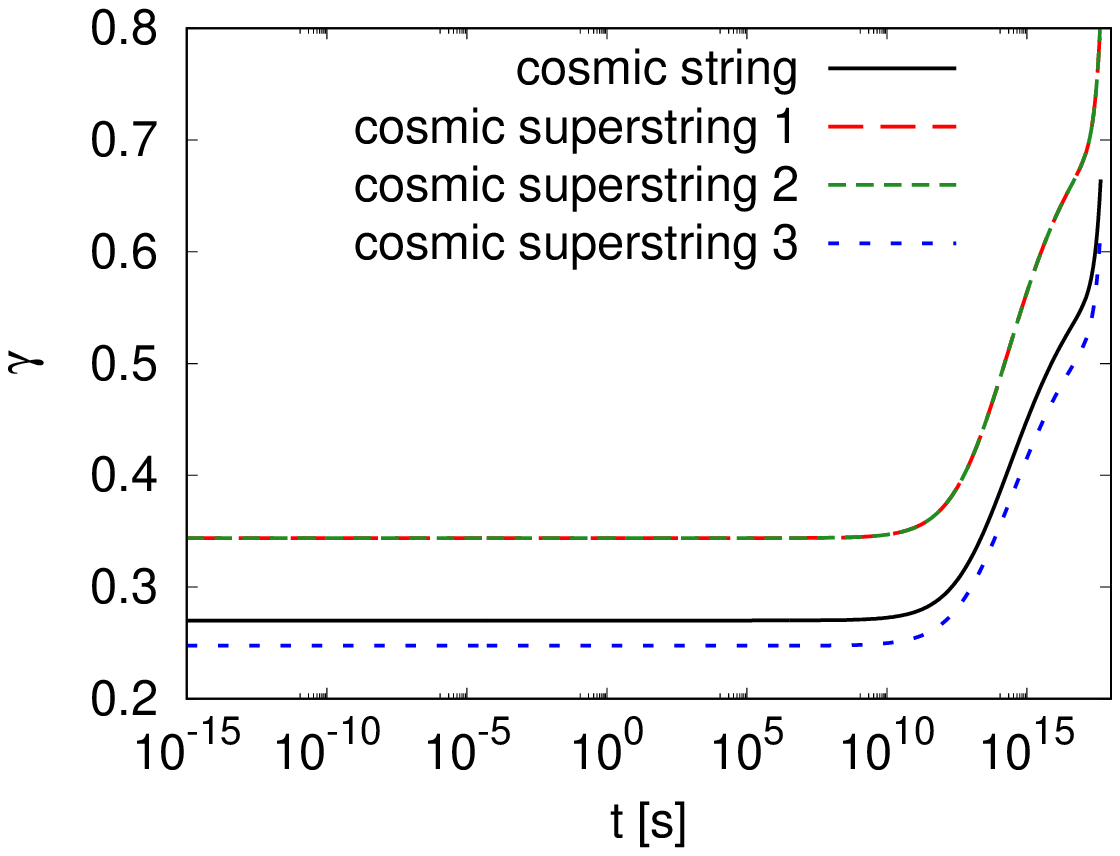}
    \end{minipage}
    \begin{minipage}{0.5\hsize}
      \centering
      \includegraphics[width=7.5cm,clip]{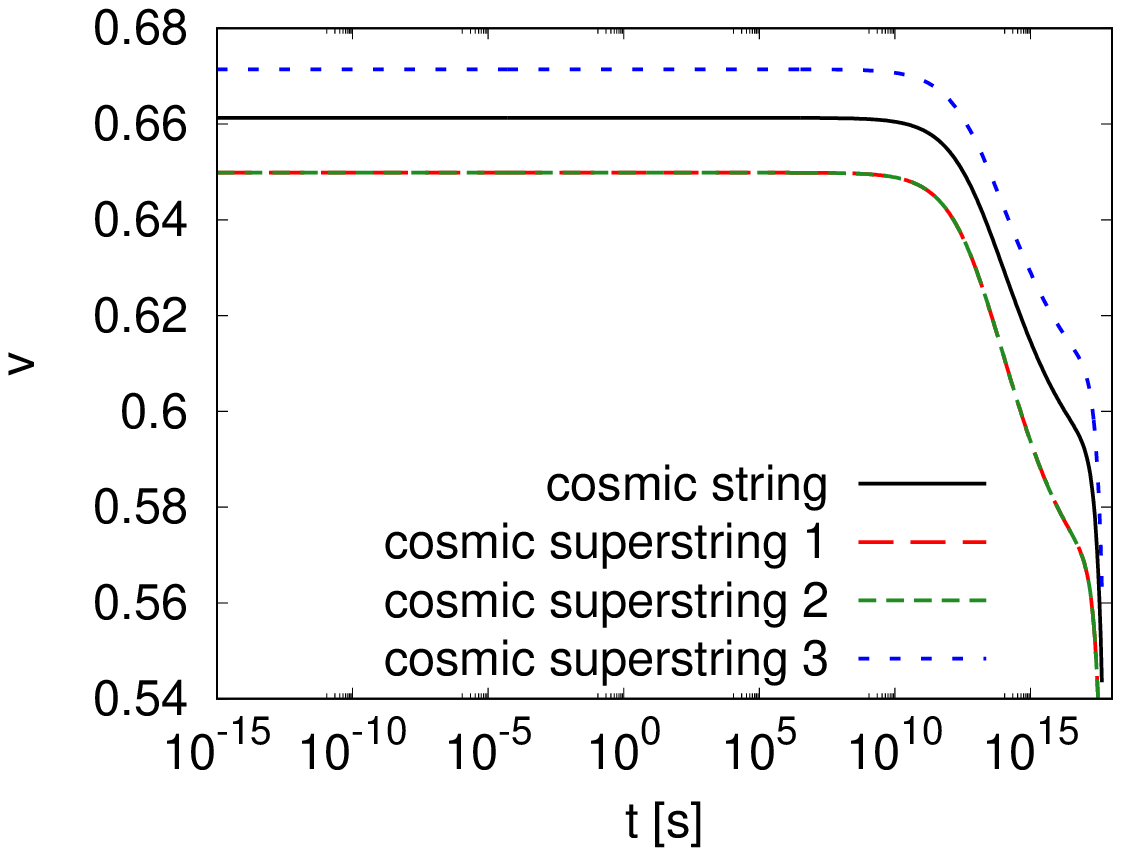}
    \end{minipage} \vspace{20pt} \\
    Case C: $\mu_1:\mu_2:\mu_3=1:10:10$, $n_p=\frac{1}{3}$, $G\mu_1=10^{-12}, \, G\mu_2=G\mu_3=10^{-11}$ \\
    \begin{minipage}{0.5\hsize}
      \centering
      \includegraphics[width=7.5cm,clip]{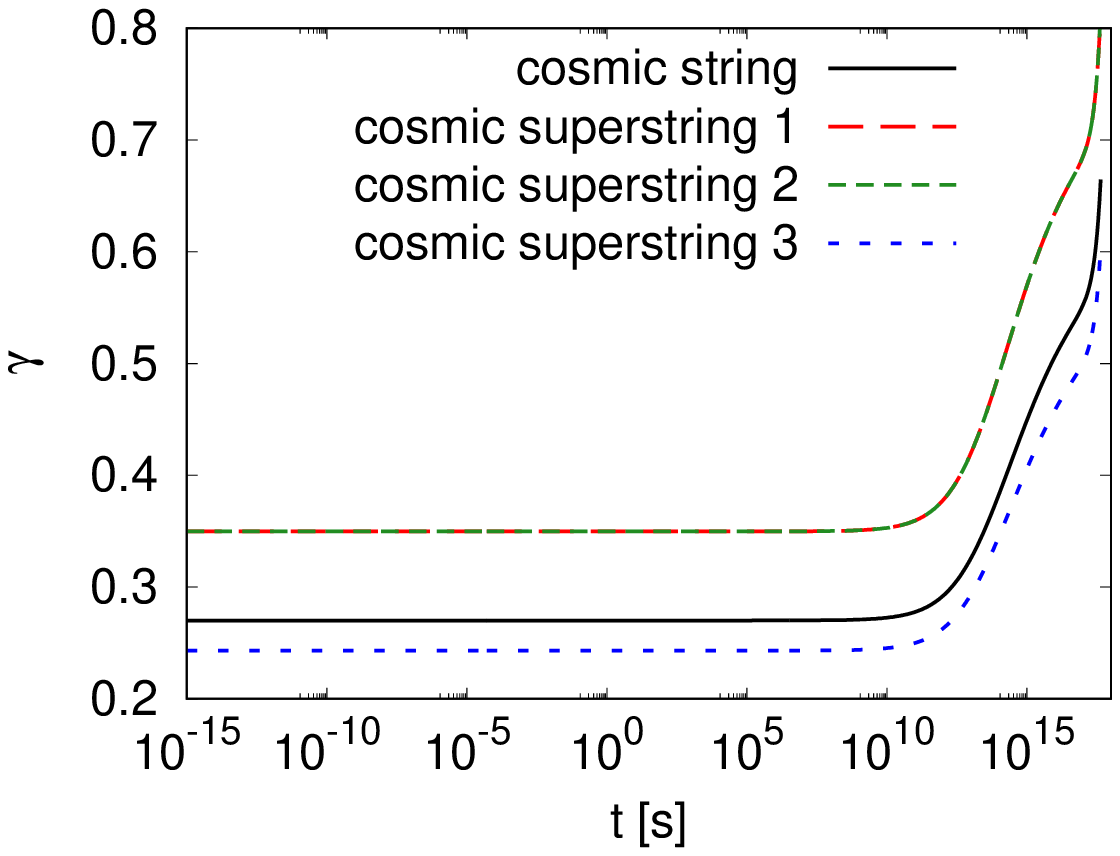}
    \end{minipage}
    \begin{minipage}{0.5\hsize}
      \centering
      \includegraphics[width=7.5cm,clip]{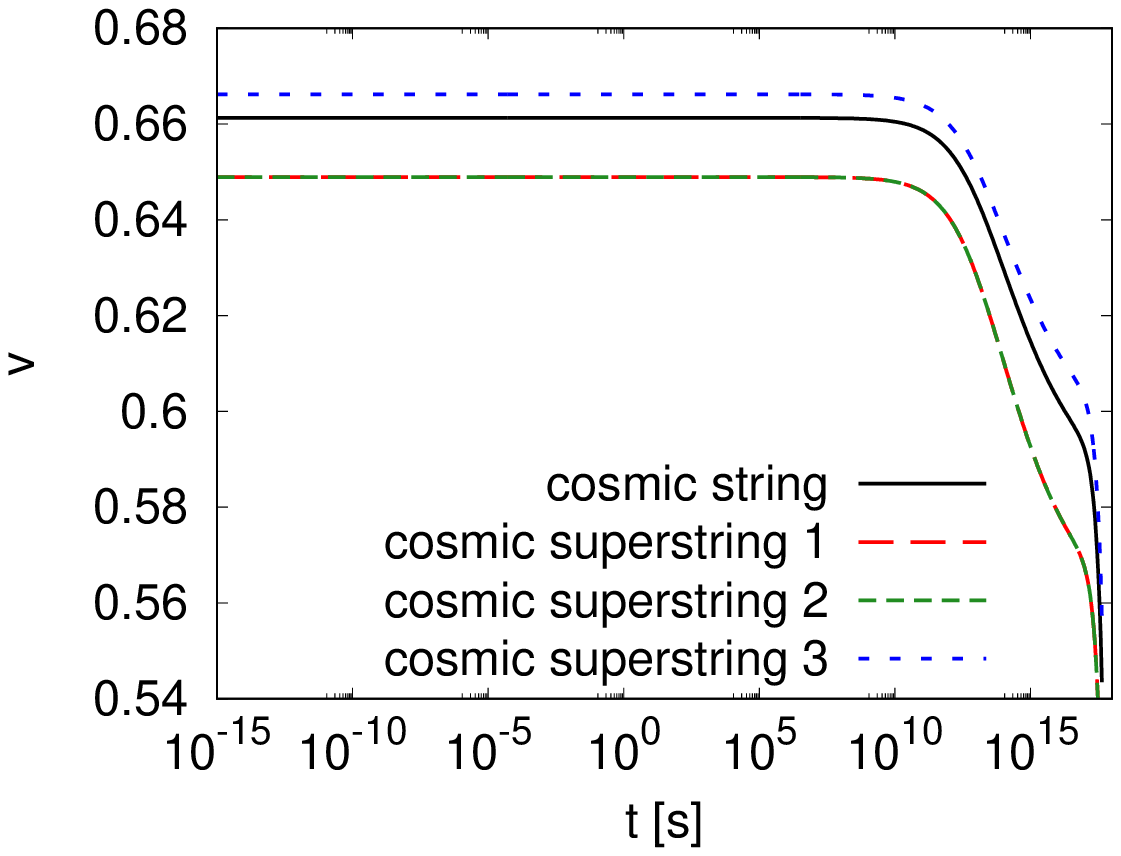}
    \end{minipage}
  \end{tabular}
  \caption{The left and the right panels respectively show the evolution of the correlation lengths and average velocities of cosmic superstrings for different string types.
The solid black line shows the case of ordinary cosmic strings presented in Sec. \ref{sec:network_cosmicstring}. The red, green and blue broken lines correspond to the type of cosmic superstrings labeled 1, 2, and 3, respectively. 
The axes are same as in Fig.~\ref{fig:scaling_CS}. The top, middle, and bottom panels show different choice of $n_p$ and string tensions, which correspond to Case A, B and C in the text. Here, the reconnection probability is fixed as $p_1=p_2=p_3=1$. The tension is assumed to be $G\mu=10^{-11}$ ($G\mu_1=10^{-12}$ and $G\mu_2=G\mu_3=10^{-11}$ for Case C).
  }
\label{fig:scaling_CSS_p=1}
\end{figure}

\begin{figure}[htbp]
  \centering
  \begin{tabular}{c}
  Case A: $\mu_1:\mu_2:\mu_3=1:1:1$, $n_p=1$, $G\mu_1=G\mu_2=G\mu_3=10^{-11}$ \\
    \begin{minipage}{0.5\hsize}
      \centering
      \includegraphics[width=7.5cm,clip]{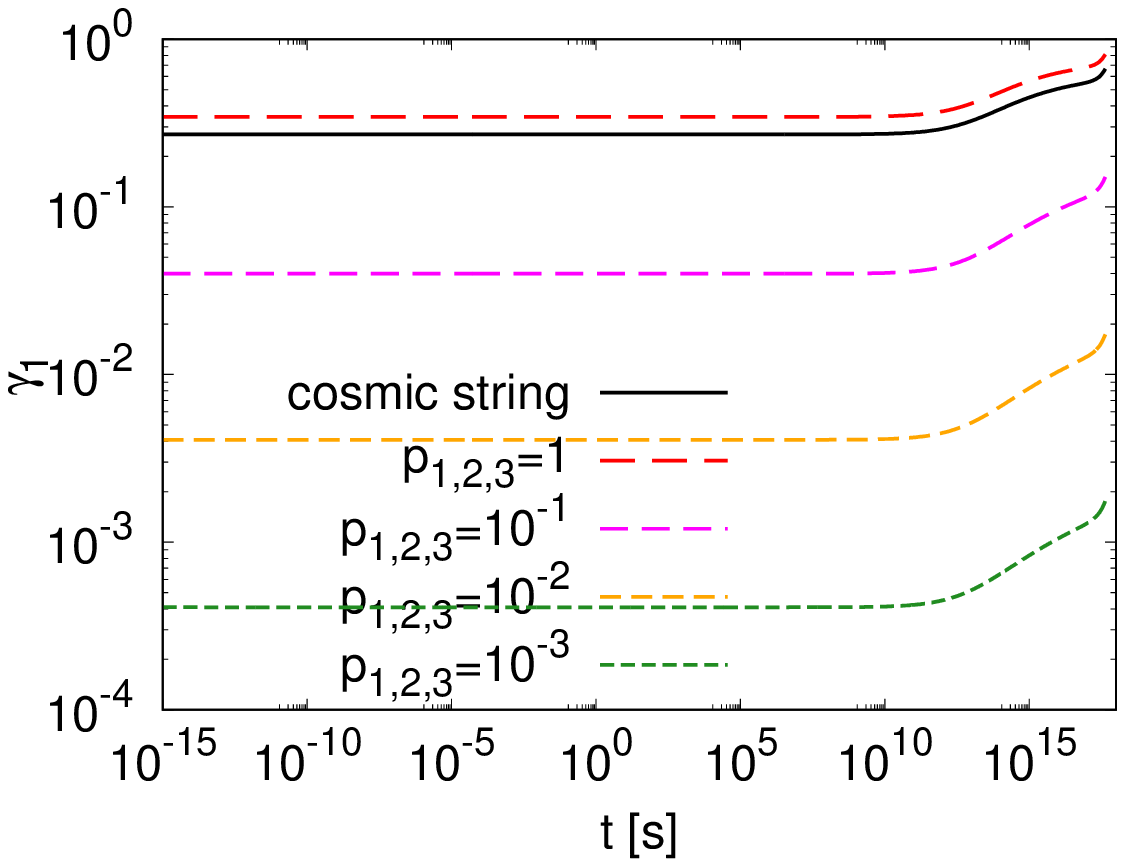}
    \end{minipage}
    \begin{minipage}{0.5\hsize}
      \centering
      \includegraphics[width=7.5cm,clip]{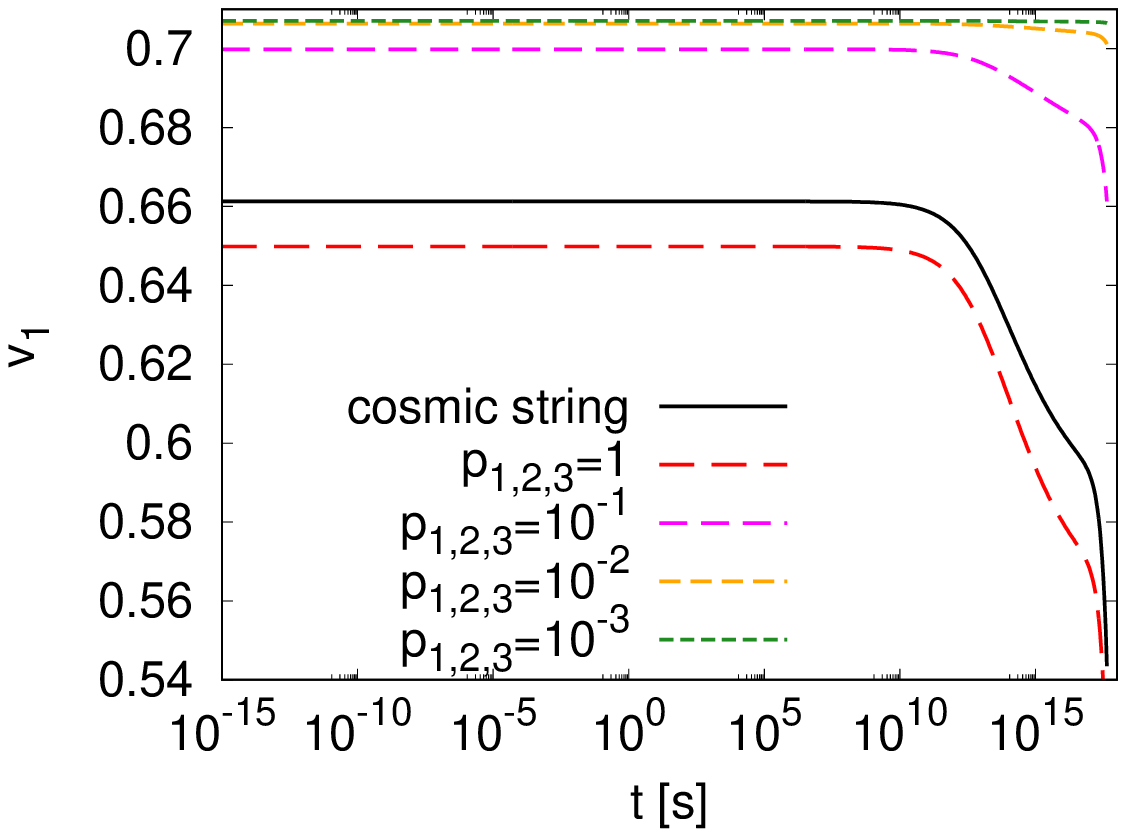}
    \end{minipage} \vspace{20pt} \\
    Case B: $\mu_1:\mu_2:\mu_3=1:1:1$, $n_p=\frac{1}{3}$, $G\mu_1=G\mu_2=G\mu_3=10^{-11}$ \\
    \begin{minipage}{0.5\hsize}
      \centering
      \includegraphics[width=7.5cm,clip]{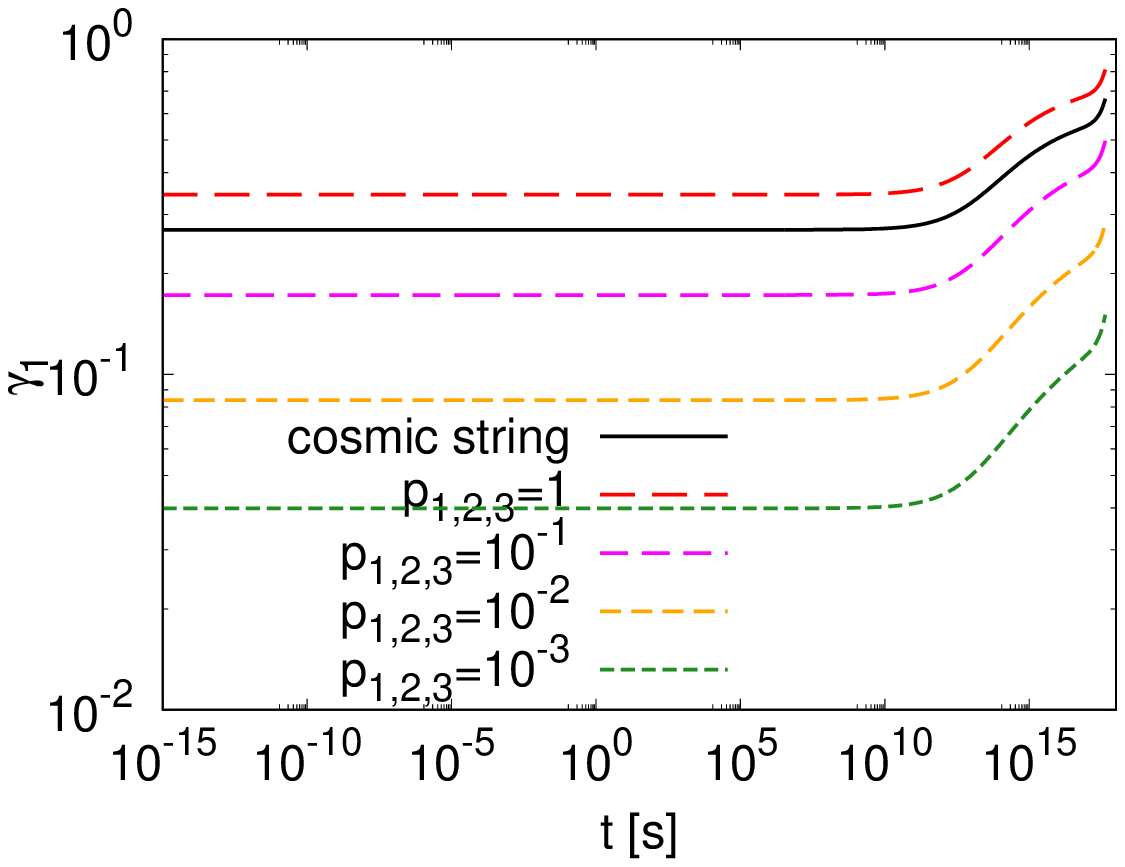}
    \end{minipage}
    \begin{minipage}{0.5\hsize}
      \centering
      \includegraphics[width=7.5cm,clip]{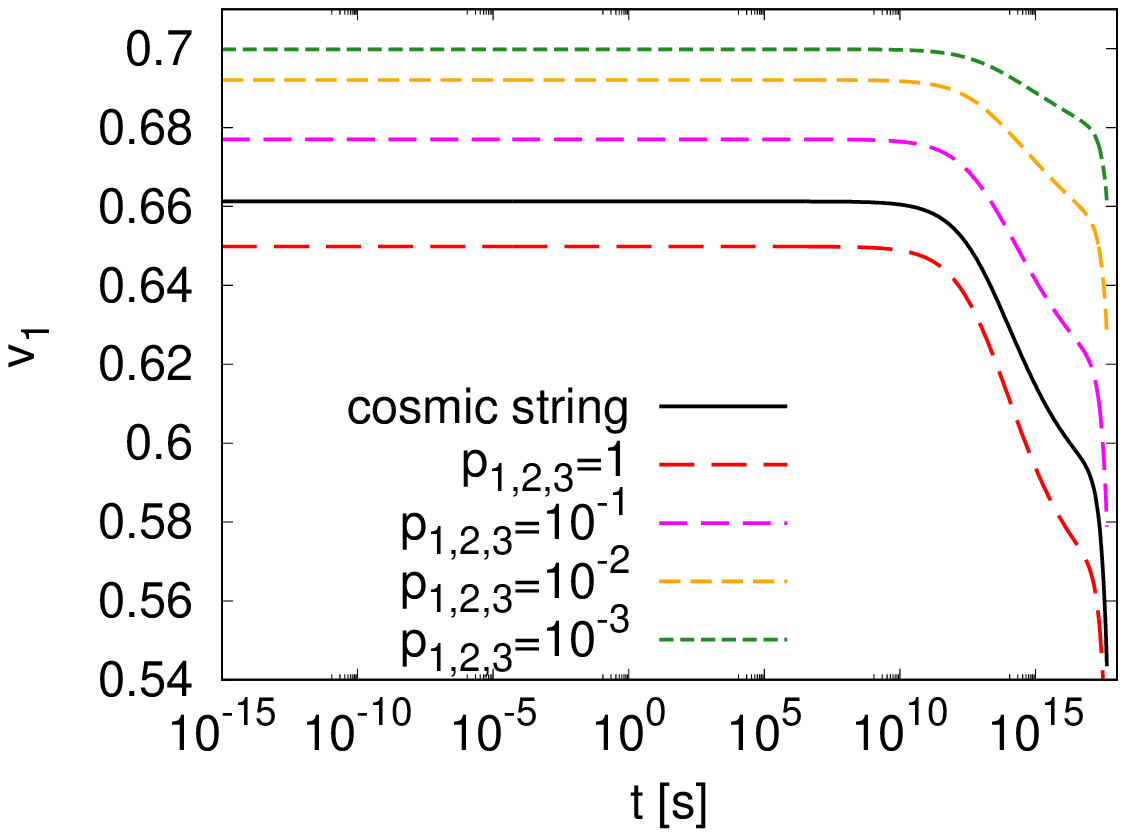}
    \end{minipage} \vspace{20pt} \\
    Case C: $\mu_1:\mu_2:\mu_3=1:10:10$, $n_p=\frac{1}{3}$, $G\mu_1=10^{-12}, \, G\mu_2=G\mu_3=10^{-11}$ \\
     \begin{minipage}{0.5\hsize}
      \centering
      \includegraphics[width=7.5cm,clip]{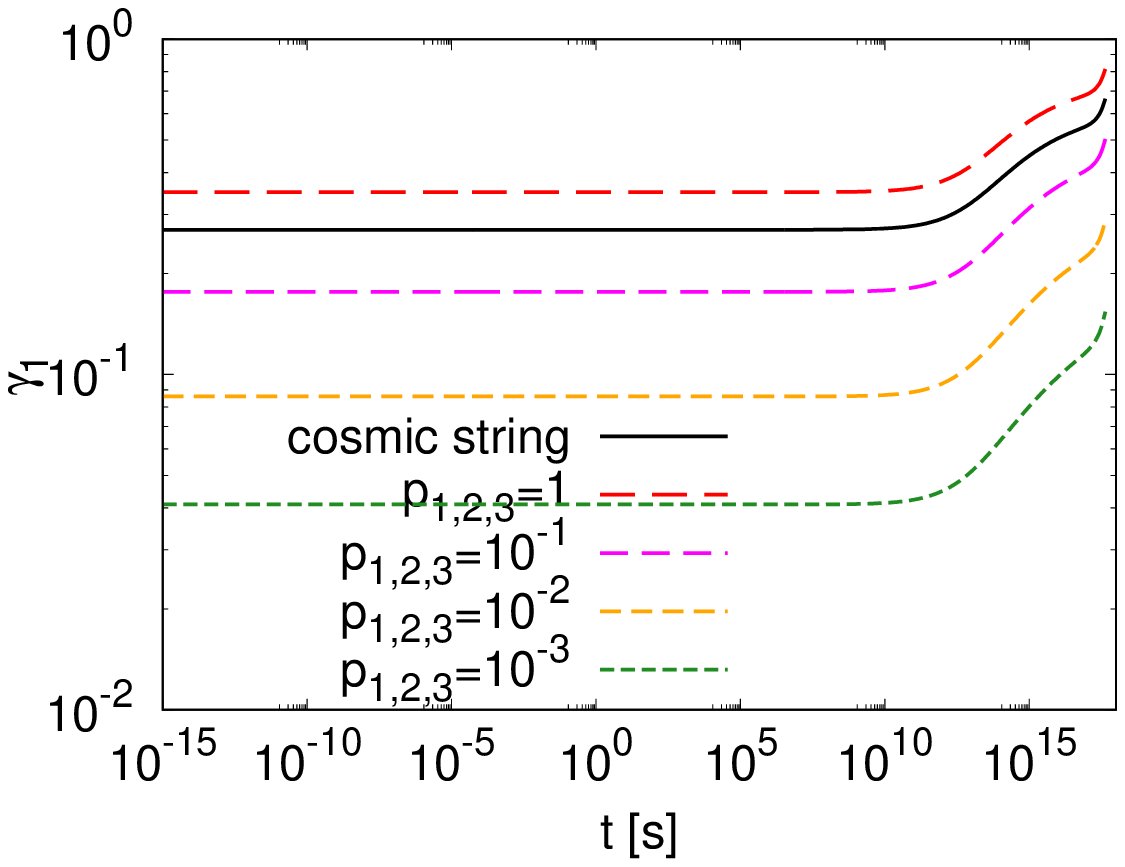}
    \end{minipage}
    \begin{minipage}{0.5\hsize}
      \centering
      \includegraphics[width=7.5cm,clip]{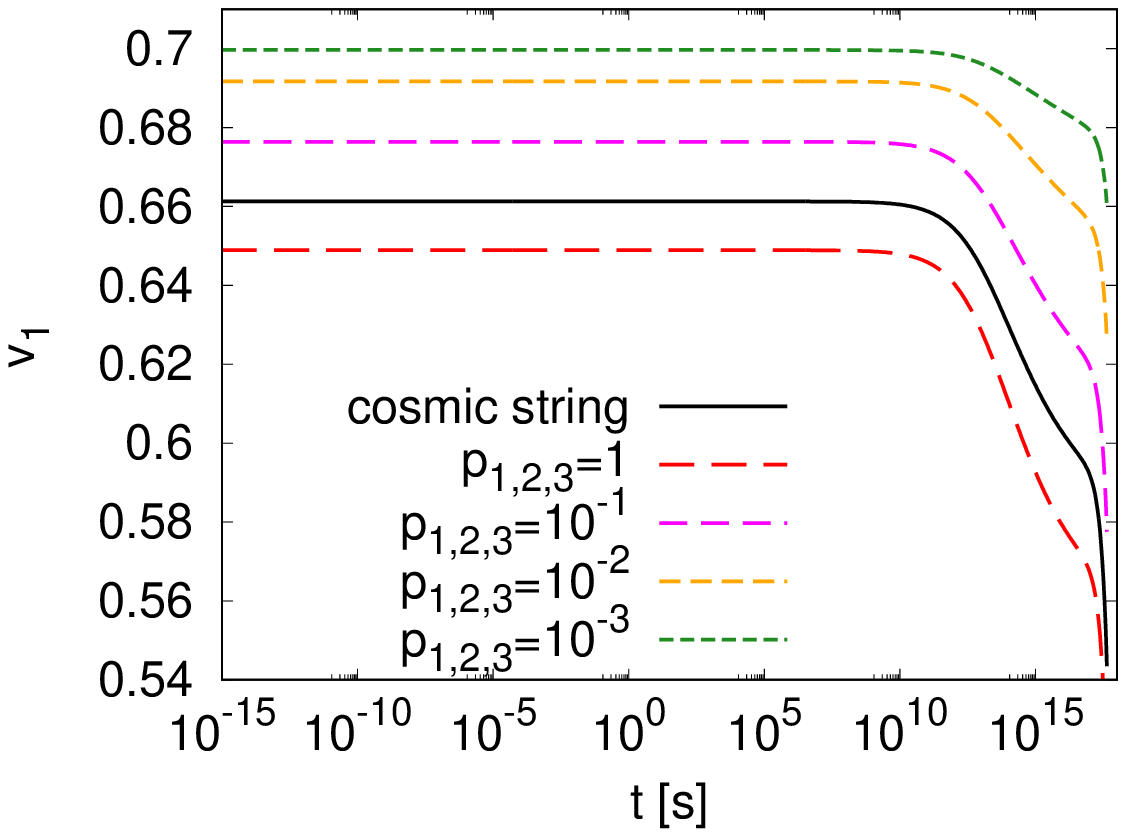}
    \end{minipage} \vspace{20pt}
  \end{tabular}
  \caption{The same plot as in Fig. \ref{fig:scaling_CSS_p=1}, but for different reconnection probabilities (only the results of string type 1 is shown).
The solid black line shows the case of ordinary cosmic strings presented in Sec. \ref{sec:network_cosmicstring}. The red, magenta, orange and green broken lines represent the cases of $p_1=p_2=p_3=1, \, 10^{-1}, \, 10^{-2}, \, 10^{-3}$, respectively.}
    \label{fig:scaling_CSS_different_p}
\end{figure}

In Figures \ref{fig:scaling_CSS_p=1} and \ref{fig:scaling_CSS_different_p}, we show the results of numerical calculations for the evolution of the correlation lengths and average velocities, obtained by simultaneously solving Eqs. \eqref{eq:L_1_CSS_eq} -- \eqref{eq:v_3_CSS_eq}. In the case of cosmic superstrings, there is a large variety of parameter choice and it is difficult to present results of all the possible parameter space. Thus, we choose three example cases for different values of string tensions and $n_p$, as a demonstration to obtain a rough idea of parameter dependence, which is listed below. We set that all string types have the same reconnection probability $p_1=p_2=p_3$ for simplicity. In Fig. \ref{fig:scaling_CSS_p=1}, we show the results for $p_1=p_2=p_3=1$, and Fig. \ref{fig:scaling_CSS_different_p} shows the results for smaller reconnection probabilities.
In both figures, the dimensionless tension of the heaviest string is set to be $G\mu=10^{-11}$, where $G$ is the gravitational constant. 

\subsubsection{Case A: string network with $\mu_1:\mu_2:\mu_3=1:1:1$ and $n_p=1$} \label{subsubsec:mu:mu:mu_np=1}
First, we consider the case where strings of type 1, 2, and 3 all have the same tension $\mu$ with low loop production efficiency $n_p=1$. 
In left top panel of Fig.~\ref{fig:scaling_CSS_p=1}, we find that ${\gamma}_1$ and ${\gamma}_2$ are larger than ${\gamma}_3$, because the third terms of Eqs.~\eqref{eq:L_1_CSS_eq} and \eqref{eq:L_2_CSS_eq}, which describes the formation of Y-junctions, increases ${\gamma}_1$ and ${\gamma}_2$, while the third term of Eq.~\eqref{eq:L_3_CSS_eq} makes ${\gamma}_3$ small. In the right top panel, we find $v_1$ and $v_2$ are slower and $v_3$ is faster than the case of ordinary cosmic strings. This is because the acceleration term of Eqs.~\eqref{eq:v_1_CSS_eq} and \eqref{eq:v_2_CSS_eq} ($k(v_1)/L_1$ and $k(v_2)/L_2$) becomes smaller due to the larger values of ${\gamma}_1$ and ${\gamma}_2$, while $k(v_3)/L_3$ becomes larger. 

In Fig. \ref{fig:scaling_CSS_different_p}, we find that $\gamma$ becomes smaller for smaller $p$.
This is because the string network with a small reconnection probability cannot produce loops efficiently, and accumulates more infinite strings inside the horizon until it can sufficiently produce loops and reaches the scaling solution.  
From our numerical result, we find the relation between the correlation length $\gamma$ and the reconnection probability $p$ in the RD era is given by 
\begin{equation}
 \gamma \propto p^{0.98}. 
\end{equation}
When the terms of Y-junction formation are absent, the relation is $\gamma \propto p$ \cite{Kuroyanagi:2012wm}.

\subsubsection{Case B: string network with $\mu_1:\mu_2:\mu_3=1:1:1$ and $n_p=\frac{1}{3}$} \label{subsubsec:mu:mu:mu_np=1over3}
Next, we consider the case where strings of type 1, 2, and 3 all have the same tension $\mu$ with high loop production efficiency $n_p=1/3$. 
The results in Fig.~\ref{fig:scaling_CSS_p=1} are the same as Case A, since the loop production term is multiplied by $p^{n_p}$ and the result does not depends on the value of $n_p$ when $p=1$.

On the other hand, in Fig.~\ref{fig:scaling_CSS_different_p}, we find the asymptotic values of the scaling solution is different from Case A when the reconnection probability is smaller than $1$. In particular, the values of $\gamma$ are larger because the loop production efficiency, determined by $n_p$, is higher in Case B and the number of infinite strings inside the horizon is reduced.
We find the relation between $\gamma$ and $p$ in the RD era is given by
\begin{equation}
 \gamma \propto p^{0.32}.
\end{equation}

\subsubsection{Case C: string network with $\mu_1:\mu_2:\mu_3=1:10:10$ and $n_p=\frac{1}{3}$} \label{subsubsec:mu:10mu:10mu_np=1over3}
Lastly, we investigate the case where the D- and F-strings have different tension. Here, we consider the case of $\mu_1:\mu_2:\mu_3=1:10:10$ with $n_p=\frac{1}{3}$. 
We do not find any remarkable difference compared to Cases A and B in Fig.~\ref{fig:scaling_CSS_p=1}, because the effect of string tension arises only in the terms of $\tilde{d}_{12}^{3}$ and the difference is small.

From Fig.~\ref{fig:scaling_CSS_different_p}, we find the asymptotic values of the scaling solution is almost the same as Case B and the dependence in the RD era is given by
\begin{equation}
 \gamma \propto p^{0.32}.
\end{equation}

\section{Distribution function of kinks} \label{sec:kink_distribution}
\subsection{The case of cosmic strings}
Let us describe how to calculate the number distribution of kinks on infinite strings. Here we first review the case of ordinary cosmic strings (namely, the network without Y-junctions and the reconnection probability $p=1$).

Kinks are continuously produced by intersections of infinite strings and their sharpness decreases with time by the expansion of the Universe.
We compute the evolution of the distribution function of kinks $N(\psi,t)d\psi$, which describes the number of kinks between $\psi$ and $\psi+d\psi$ within the arbitrary volume $V$. The evolution equation of the distribution function is given by 
\cite{Copeland:2009dk,Kawasaki:2010yi,Matsui:2016xnp}
\begin{equation}
    \frac{\partial N}{\partial t}(\psi, \, t) = \frac{\bar{\Delta} V}{\gamma^4 t^4} g(\psi) +\frac{2 \zeta}{t} \frac{\partial}{\partial \psi} (\psi N(\psi, \, t)) -\frac{\eta}{\gamma t} N(\psi, \, t),  \label{eq:kink_eq}
\end{equation}
with the parameters
\begin{eqnarray}
  \bar{\Delta}
  & = & \frac{2 \pi}{35} \Bigl\{1 +\frac{2}{3}(1-2v^2) -\frac{1}{11}(1 -2v^2)^2 \Bigr\}, \\
  \zeta & = & (1 -2v^2) \nu,  \\
  \eta & = & \frac{1}{2} c p^{n_p} v.
\end{eqnarray}
The first term on the right hand side describes the kink production. The parameter $\bar{\Delta}$ characterizes the
rate of intersections of infinite strings \cite{Austin:1993rg}, and $g(\psi)$ gives the initial sharpness distribution, $g(\psi) = \frac{35}{256} \sqrt{\psi}(15 -6\psi-\psi^2)$ for $0 \leqq \psi \leqq 1$ \cite{Copeland:2009dk}.
The second term gives the effect of decreasing kink sharpness due to the expansion of the Universe, as described in Eq. \eqref{eq:zeta}. The third term gives the loss of kinks going to loops.

\begin{figure}[htbp]
\centering
\includegraphics[width=12cm,clip]{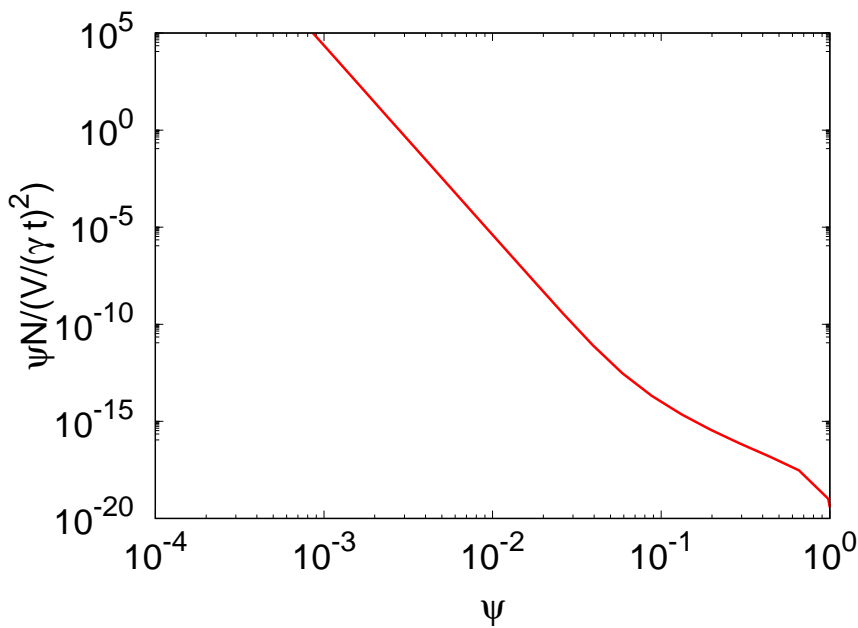}
\caption{The distribution function of kinks on infinite strings. The horizontal axis is the sharpness of kinks and the vertical axis is the the number of kinks per unit length of infinite strings.}
\label{fig:kink_distribution_CS}
\end{figure}

By numerically solving the evolution equation \eqref{eq:kink_eq}, we obtain the distribution function of kinks, which is shown in Fig.~\ref{fig:kink_distribution_CS}. Since kinks are blunted by the expansion of the Universe, older kinks have smaller sharpness. Thus, in Fig.~\ref{fig:kink_distribution_CS}, kinks in the left side (with small sharpness) are produced in the RD era, and kinks on the right side (with large sharpness) are produced in the MD era.

\subsection{The case of cosmic superstrings}
Cosmic superstrings have Y-junctions in the network, where D- and F-strings meet and form bound states. When a kink enters a Y-junction, two transmitted kinks and one reflected kink appear and the sharpness of these kinks differs from the original incoming kink depending on the tension and the angles of the three strings. Here, we define the transmission coefficient 
\begin{equation}
  C_{ij}=\frac{\psi_j^{(\rm out)}}{\psi_i^{(\rm in)}},
\end{equation}
where $i,j=1,2,3$ label different strings connecting to the Y-junction, and ``in'' and ``out'' denote incoming and outgoing kinks. 
For example, $C_{12}$ describes the ratio of the sharpness of the outgoing kink on string 2 to the incoming kink from string 1, while $C_{11}$ gives the sharpness of the kink reflected onto string 1.
Reference \cite{Binetruy:2010bq} made a detailed study on how the sharpness of the incoming kink is transmitted to the three daughter kinks using numerical simulations. Although the transmission coefficient was found to be distributed over a wide range of values depending on the configuration of the three strings, we adopt the average value of the coefficient. As far as the choice of tensions is concerned, we investigate two cases: $\mu_1:\mu_2:\mu_3=1:1:1$ and $\mu_1:\mu_2:\mu_3=1:10:10$.

\begin{figure}[htbp]
\centering
\includegraphics[width=12cm,clip]{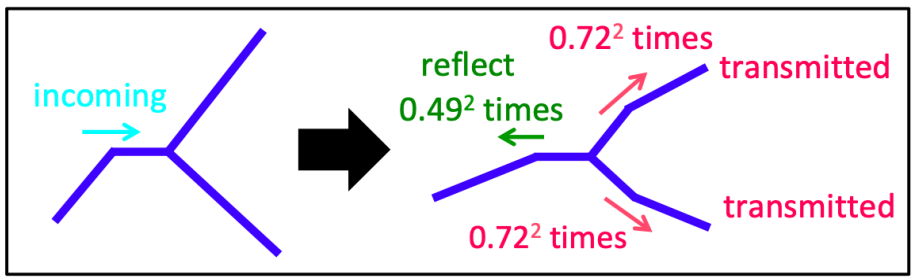}
\caption{The figure illustrates how the sharpness of a kink is altered when it goes through a Y-junction for the case of $\mu_1:\mu_2:\mu_3=1:1:1$.}
\label{fig:Y-junction_change_sharpness_at_mu:mu:mu}
\end{figure}

In the case of equal tensions, the sharpness of the reflected and transmitted kinks are $0.49^2$ and $0.72^2$ times smaller than the incoming kink in average, respectively (the value is taken from Fig. 3 of \cite{Binetruy:2010bq}).
\footnote{Note that the definition of the sharpness in \cite{Binetruy:2010bq} is $|\sin(\theta/2)|$, where $\theta$ is the kink angle, while our definition of $\psi$, Eq.~\eqref{eq:psi_define}, is transformed to $\sin^2(\theta/2)$ \cite{Copeland:2009dk}. Thus, the values of the transmission coefficient obtained in \cite{Binetruy:2010bq} are squared in this paper. 
}
The picture is given in Fig.~\ref{fig:Y-junction_change_sharpness_at_mu:mu:mu}. In summary, the transmission coefficient is given by 
\begin{equation}
  C_{ij}=
\left(
\begin{array}{rrr}
    C_{11} & C_{12} & C_{13} \\ 
    C_{21} & C_{22} & C_{23} \\ 
    C_{31} & C_{32} & C_{33} \\ 
\end{array}
\right)
=
\left(
\begin{array}{rrr}
    0.49^2 & ~ 0.72^2 & ~ 0.72^2 \\ 
    0.72^2 & 0.49^2 & 0.72^2 \\ 
    0.72^2 & 0.72^2 & 0.49^2 \\ 
\end{array}
\right).
\label{eq:C_ij_mu:mu:mu}
\end{equation}

\begin{figure}[htbp]
\centering
\includegraphics[width=12cm,clip]{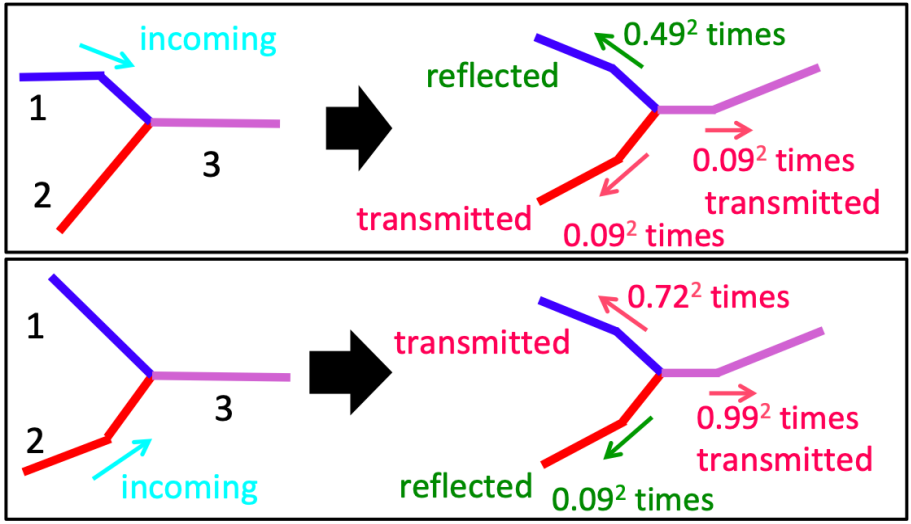}
\caption{The figure illustrates how the sharpness of a kink is altered when it goes through a Y-junction for the case of $\mu_1:\mu_2:\mu_3=1:10:10$.
}
\label{fig:Y-junction_change_sharpness_at_mu:10mu:10mu}
\end{figure}

We also consider non-equal tensions with the ratio of $\mu_1:\mu_2:\mu_3=1:10:10$. When the incoming kink comes from the light string, the sharpness of the kinks transmitted to the other two heavy strings is $0.09^2$ times smaller, while the reflected kink becomes $0.49^2$ times. When the incoming kink is from the heavy string, the sharpness changes depending on the tension of the transmitted strings. The transmission coefficient is $0.72^2$ for the light string and $0.99^2$ for the heavy string. The reflected kink becomes $0.09^2$ times smaller. See Appendix A.1 and A.2 of \cite{Binetruy:2010bq} for the details. The values are summarized in Fig.~\ref{fig:Y-junction_change_sharpness_at_mu:10mu:10mu} and the transmission coefficient is given by
\begin{equation}
  C_{ij}=
\left(
\begin{array}{rrr}
  0.49^2 & ~0.09^2 & ~0.09^2 \\ 
  0.72^2 & 0.09^2 & 0.99^2 \\ 
  0.72^2 & 0.99^2 & 0.09^2 \\
\end{array}
\right).
\label{eq:C_ij_mu:10mu:10mu}
\end{equation}

Using the transmission coefficient, we find the evolution equations of the kink distribution function for string types $1$, $2$ and $3$ are given, respectively, as 
\begin{eqnarray}
  \frac{\partial N_1}{\partial t}(\psi, \, t) & = & \frac{p_1\bar{\Delta_1} V}{\gamma_1^4 t^4} g(\psi) +\frac{2 \zeta_1}{t} \frac{\partial}{\partial \psi} (\psi N_1(\psi, \, t)) -\frac{\eta_1}{\gamma_1 t} N_1(\psi, \, t) \nonumber  \\ 
 & \, & \hspace{5pt} +\frac{\alpha}{t} N_2 \left(\frac{\psi}{C_{21}}, \, t \right ) +\frac{\alpha}{t} N_3 \left(\frac{\psi}{C_{31}}, \, t \right ) +\frac{\alpha}{t} N_1 \left(\frac{\psi}{C_{11}}, \, t \right ) -\frac{\alpha}{t} N_1(\psi, \, t),  \label{eq:kink_eq_CSS_1}  \\
   \frac{\partial N_2}{\partial t}(\psi, \, t) & = & \frac{p_2\bar{\Delta_2} V}{\gamma_2^4 t^4} g(\psi) +\frac{2 \zeta_2}{t} \frac{\partial}{\partial \psi} (\psi N_2(\psi, \, t)) -\frac{\eta_2}{\gamma_2 t} N_2(\psi, \, t) \nonumber  \\ 
 & \, & \hspace{5pt} +\frac{\alpha}{t} N_1 \left(\frac{\psi}{C_{12}}, \, t \right ) +\frac{\alpha}{t} N_3 \left(\frac{\psi}{C_{32}}, \, t \right ) +\frac{\alpha}{t} N_2 \left(\frac{\psi}{C_{22}}, \, t \right ) -\frac{\alpha}{t} N_2(\psi, \, t),  \label{eq:kink_eq_CSS_2}  \\
    \frac{\partial N_3}{\partial t}(\psi, \, t) & = & \frac{p_3\bar{\Delta_3} V}{\gamma_3^4 t^4} g(\psi) +\frac{2 \zeta_3}{t} \frac{\partial}{\partial \psi} (\psi N_3(\psi, \, t)) -\frac{\eta_3}{\gamma_3 t} N_3(\psi, \, t) \nonumber  \\ 
 & \, & \hspace{5pt} +\frac{\alpha}{t} N_1 \left(\frac{\psi}{C_{13}}, \, t \right ) +\frac{\alpha}{t} N_2 \left(\frac{\psi}{C_{23}}, \, t \right ) +\frac{\alpha}{t} N_3 \left(\frac{\psi}{C_{33}}, \, t \right ) -\frac{\alpha}{t} N_3(\psi, \, t).  \label{eq:kink_eq_CSS_3}
\end{eqnarray}
The parameters $\bar{\Delta_i}, \, \zeta_i$ and $\eta_i$ are given by the average velocity of each string $v_i$ as 
\begin{eqnarray}
\bar{\Delta}_i & = & \frac{2 \pi}{35} \Bigl\{1 +\frac{2}{3}(1 -2v_i^2) -\frac{1}{11}(1 -2v_i^2)^2 \Bigr\}\\
\zeta_i & = & (1 -2v_i^2)\nu,\\
\eta_i & = & \frac{1}{2}c_ip_i^{n_p}v_i.
\end{eqnarray}

Compared to the ordinary cosmic string case described in Eq.~\eqref{eq:kink_eq}, we added two effects associating with the characteristics of the superstring network. 
First, the kink production term is multiplied by $p$ as kinks are generated by intersections of strings and the number of intersections is reduced linearly by the reconnection probability. 
Second, the new terms are added in order to include the effect of kinks entering Y-junction. The fourth and fifth terms describe kinks coming from different types of strings by changing their sharpness. The sixth term corresponds to reflected kinks. These terms describe that kinks, whose sharpness was $\psi_i^{(in)}=\psi_j^{(out)}/C_{ij}$, transmit to different or the same string with the rate of $\alpha$ times per horizon time. The seventh term describes the disappearance of incoming kinks.

For the value of $\alpha$, we use the following estimation.
Since kinks propagate with the speed of light, kinks move $\sim t$ in a Hubble time, while the average distance between Y-junction would be roughly given by the correlation length of three strings as $\sim (\gamma_1 + \gamma_2 + \gamma_3)t/3$. Thus, the number of kinks encountering Y-junctions in a Hubble time is roughly given by 
\begin{equation}
\alpha = \frac{3}{\gamma_1 + \gamma_2 + \gamma_3}.
\label{eq:alpha}
\end{equation}

In Fig. \ref{fig:kink_distribution_CSS}, we show the distribution function of kinks obtained by numerically calculating Eqs.\eqref{eq:kink_eq_CSS_1} -- \eqref{eq:kink_eq_CSS_3} and Eq.~\eqref{eq:L_1_CSS_eq} -- Eq.~\eqref{eq:v_3_CSS_eq}.  
From the top to the bottom, we show Case A, B and C. For the transmission coefficient, we use Eq.\eqref{eq:C_ij_mu:mu:mu} for Cases A and B, and Eq.\eqref{eq:C_ij_mu:10mu:10mu} for Case C. 
The left panels show results for different string types $1$, $2$ and $3$ with $p_1 = p_2 = p_3 = 1$.
In the right panels, we show how different reconnection probabilities affect the results.
The lines represent the sum of the kink number of different string types for $p_1=p_2=p_3=1, \, 10^{-1}, \, 10^{-2}$ and $10^{-3}$. For comparison, we also plot the result of ordinary cosmic strings (same as Fig.~\ref{fig:kink_distribution_CS}).

\subsubsection{Case A: kink distribution with $\mu_1:\mu_2:\mu_3=1:1:1$ and $n_p=1$} \label{subsubsec:kink_distribution_mu:mu:mu_np=1}

In the left panel, the reconnection probabilities are all set to unity, so that we can easily see the effect of Y-junctions by comparing it with the ordinary cosmic string case, which should be identical when we set $p_1 = p_2 = p_3 = 1$ and remove the Y-junction terms. In all cases, we find that the number distribution is flatter when we include the effect of Y-junctions. This is because, when a kink enters a Y-junction, three transmitted kinks typically have smaller sharpness than the original one. Thus, Y-junctions increase the number of kinks with small sharpness, while they decrease the number of kinks with large sharpness, which flattens the distribution and extends the cutoff to a much lower sharpness \footnote{Note that, although the figures may give the impression that the total number of kinks decreases when we include the effect of Y-junctions, this is not true since there are a huge number of kinks at much smaller sharpness beyond the plot range of the figure. In fact, Y-junctions increase the total number of kinks. However, the increased kinks have very small sharpness and they do not increase the GW background amplitude, as we will see in the next sections.}.

In the right panel, we find that a smaller reconnection probability tends to flatten the distribution more. This is the result of two combined effects. First, a small $p$ decreases the correlation length and increases the number of strings inside the horizon. This enhances the kink production term, as it is proportional to $p/\gamma^4$ and $\gamma \propto p$ for Case A, and increases the overall number of kinks. On the other hand, the small correlation length increases the number of kinks encountering Y-junctions, which is parametrized by $\alpha$, and makes the effect of Y-junction terms stronger. In Case A, we find that the latter effect is always stronger than the former, and we find that the distribution is more flattened for smaller $p$.

\subsubsection{Case B: kink distribution with $\mu_1:\mu_2:\mu_3=1:1:1$ and $n_p=\frac{1}{3}$} \label{subsubsec:kink_distribution_mu:mu:mu_np=1over3}
In Case B, the kink distribution looks similar to Case A. The only difference arises in the right panel, where we find the number of kinks is smaller than Case A when the reconnection probability is small. This is because the loop production is more efficient in the case of $n_p=\frac{1}{3}$, and the decrease of correlation length $\gamma$ is milder for a smaller reconnection probability as shown in the right panel of Fig.~\ref{fig:scaling_CSS_different_p} and found $\gamma\propto p^{0.32}$. Since $\gamma$ has a larger value, the kink production term $\propto p/\gamma^4$ is suppressed, which is the reason that we find less number of kinks. 

\subsubsection{Case C: kink distribution with $\mu_1:\mu_2:\mu_3=1:10:10$ and $n_p=\frac{1}{3}$} \label{subsubsec:kink_distribution_mu:10mu:10mu_np=1over3}
In the left panel of Case C, we find the slope of the distribution function is more flattened compared to the cosmic string case because of the existence of Y-junctions, but not as much as Cases A and B.
This is because kinks are smoothed out efficiently with the coefficient $0.49^2$ and $0.72^2$ in Cases A and B, while we have the coefficient of $C_{32}=C_{23}=0.99^2$ in Case C, which means that one of the three kinks stays with the original sharpness when a kink enters from type $3$ and $2$, which accounts for $2/3$ of intersections with Y-junction. Thus, the effect of Y-junctions to smooth out kinks is weaker in Case C. 
We also find the number of kinks on string $1$ is smaller than types $2$ and $3$. This is because the kinks with the original sharpness (with the coefficient of $0.99^2$) are transmitted to string $2$ or $3$, and the sharpness of kinks going to string $1$ is always multiplied by $0.49^2$ or $0.72^2$. Thus, kinks on string $1$ tend to get flat more compared to the ones for strings $2$ and $3$.

In the right panel, we find an interesting behavior that the number of kinks slightly increases in the case of $p=10^{-1}$, compared to the case of $p=1$, because the increase of the kink production by a smaller value of $\gamma$ dominates the smoothing out of kink sharpness by Y-junctions. When the value of $p$ decreases to $10^{-2}$ and $10^{-3}$, the latter effect becomes stronger and the distribution gets flat.

\begin{figure}[htbp]
  \centering
  \begin{tabular}{c}
    Case A: $\mu_1:\mu_2:\mu_3=1:1:1$, $n_p=1$, $G\mu_1=G\mu_2=G\mu_3=10^{-11}$ \\
    \begin{minipage}{0.5\hsize}
      \centering
      \includegraphics[width=7.5cm,clip]{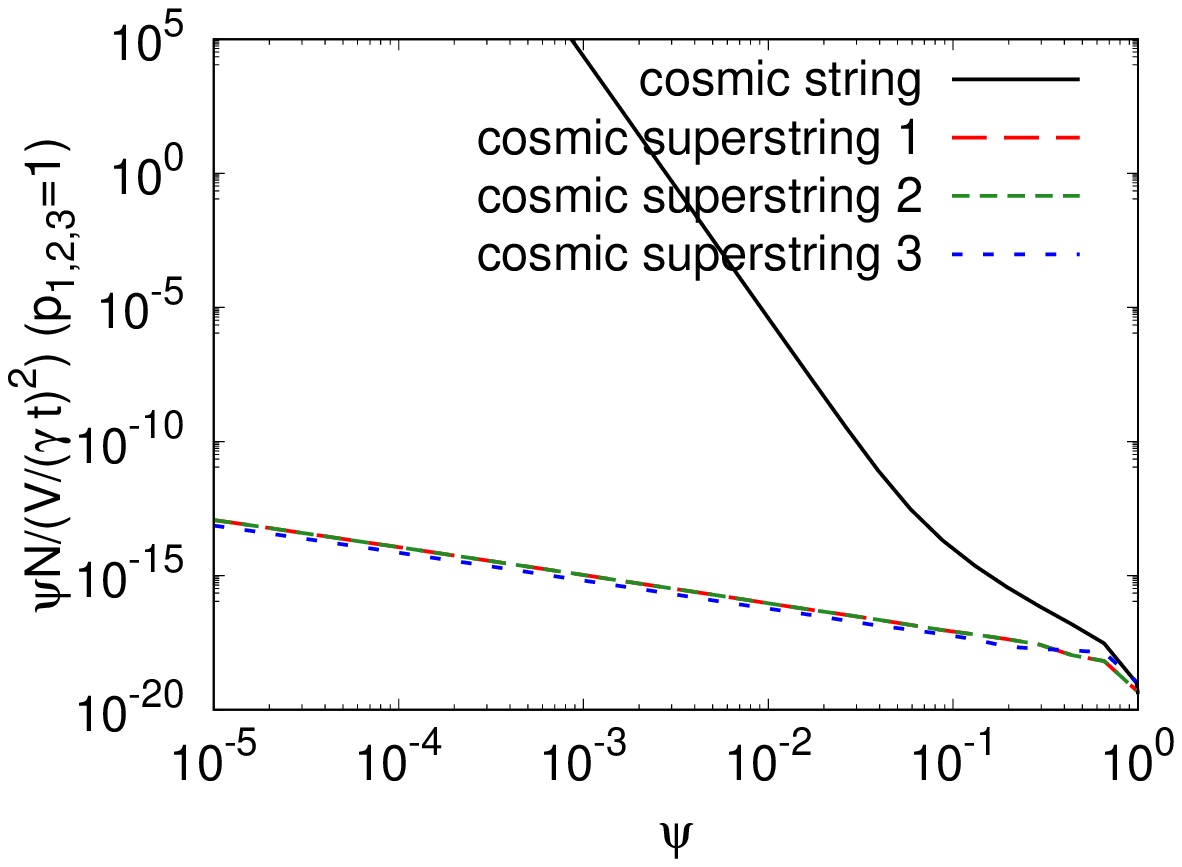}
    \end{minipage}
    \begin{minipage}{0.5\hsize}
      \centering
      \includegraphics[width=7.5cm,clip]{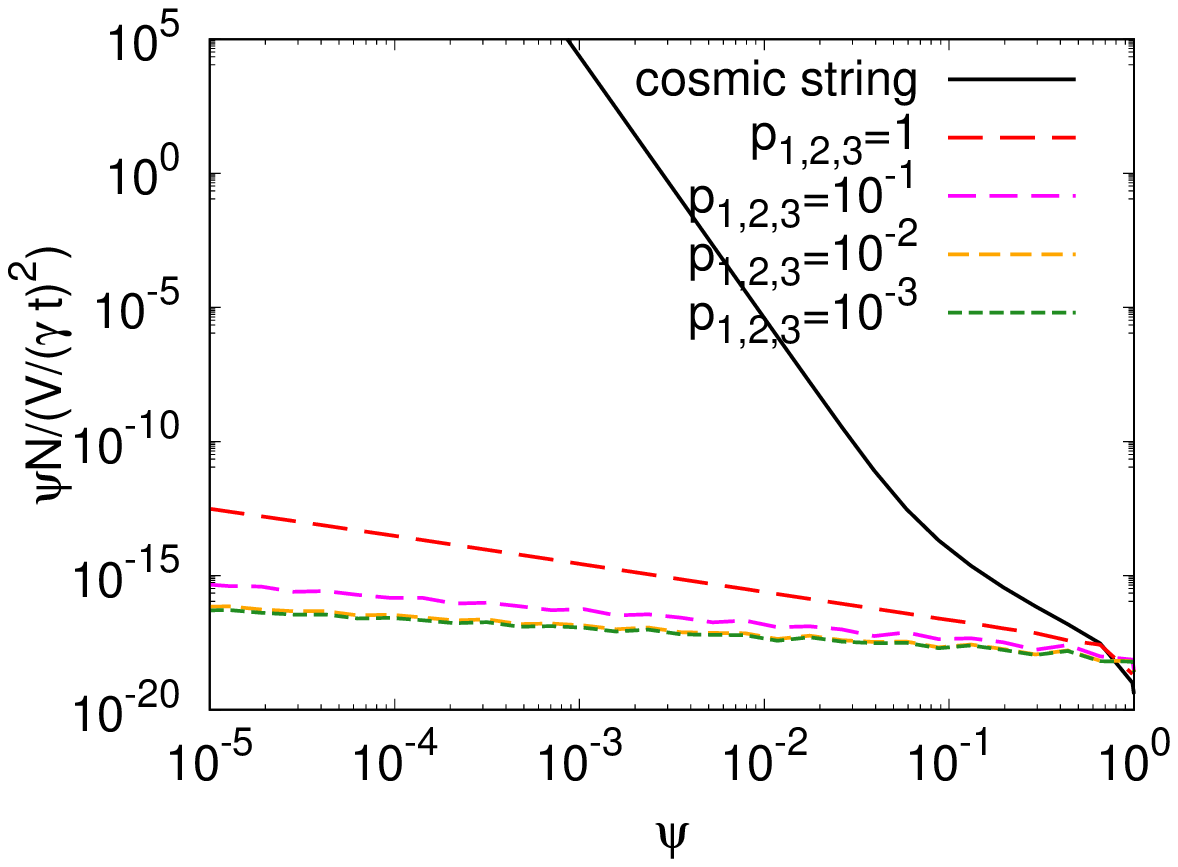}
    \end{minipage} \vspace{20pt} \\
    Case B: $\mu_1:\mu_2:\mu_3=1:1:1$, $n_p=\frac{1}{3}$, $G\mu_1=G\mu_2=G\mu_3=10^{-11}$ \\
        \begin{minipage}{0.5\hsize}
      \centering
      \includegraphics[width=7.5cm,clip]{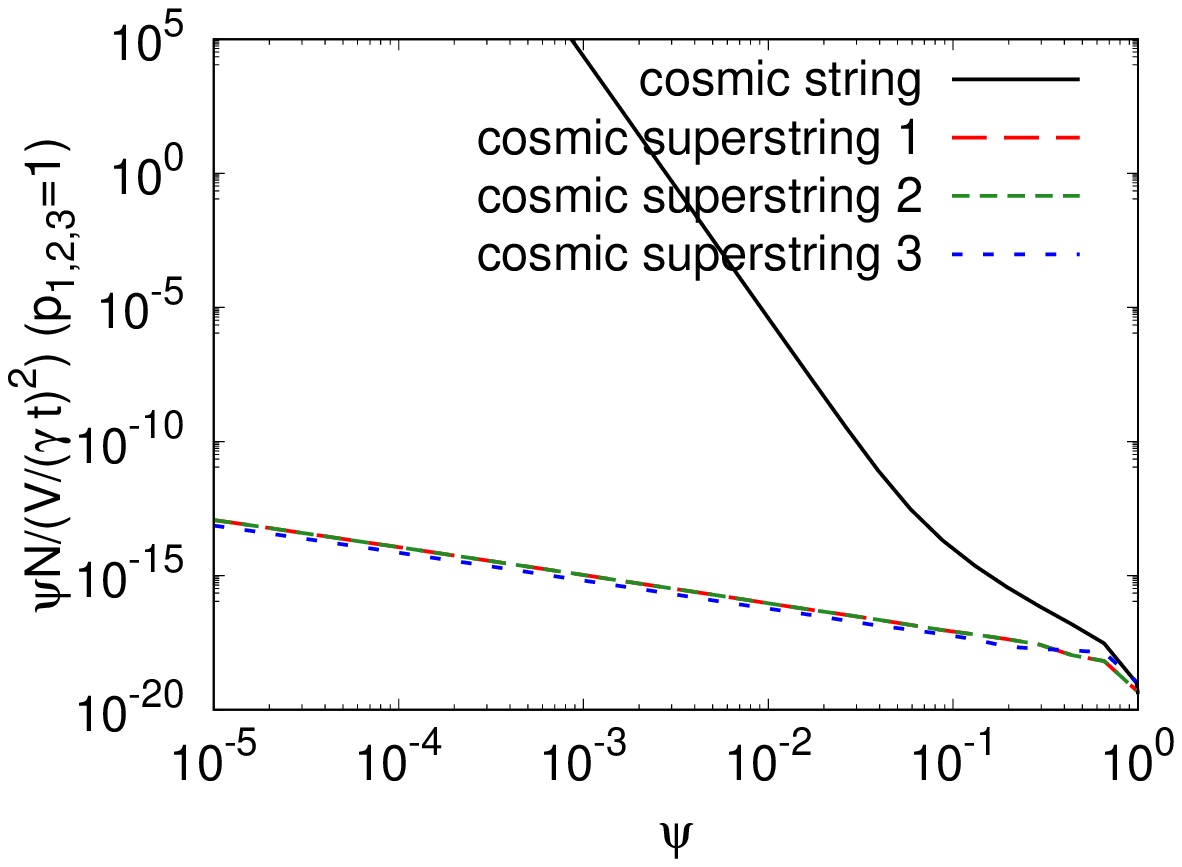}
    \end{minipage}
    \begin{minipage}{0.5\hsize}
      \centering
      \includegraphics[width=7.5cm,clip]{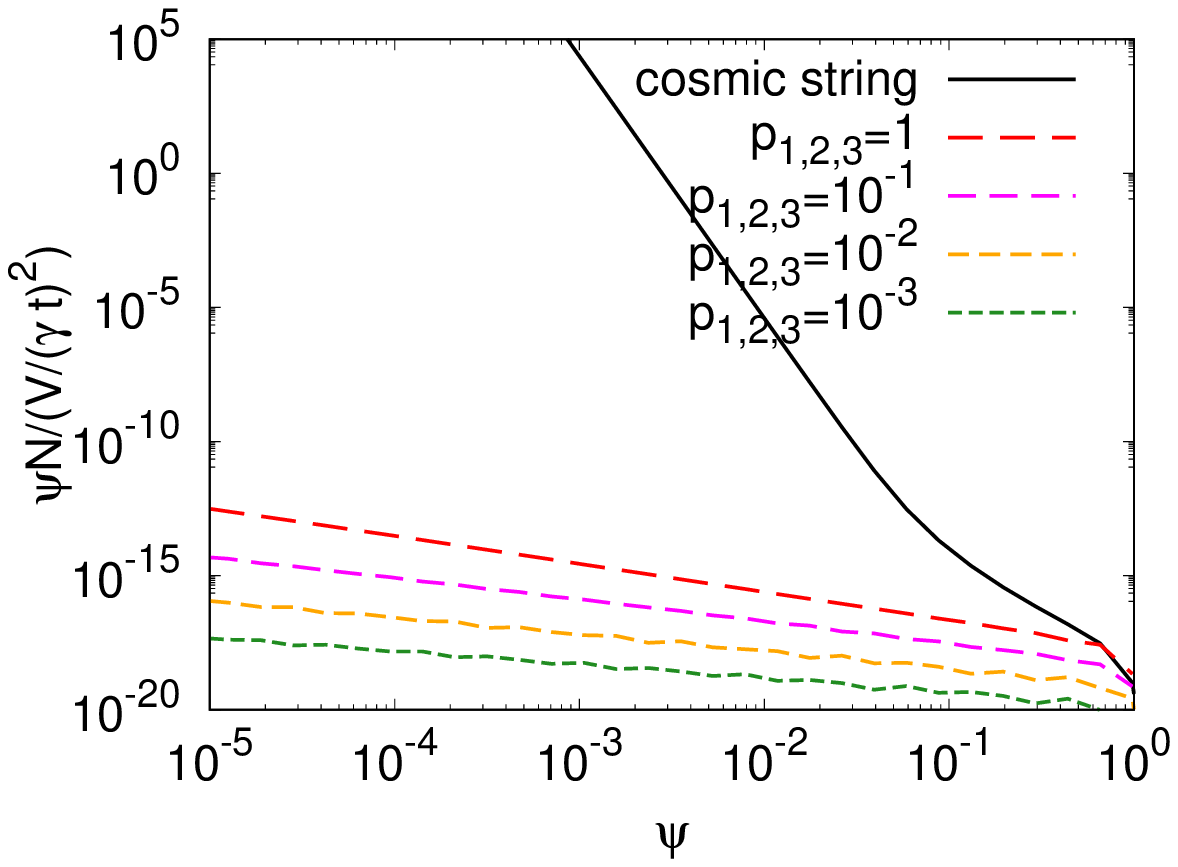}
    \end{minipage} \vspace{20pt} \\
    Case C: $\mu_1:\mu_2:\mu_3=1:10:10$, $n_p=\frac{1}{3}$, $G \mu_1 = 10^{-12}, \, G \mu_2 = G \mu_3 = 10^{-11}$ \\
    \begin{minipage}{0.5\hsize}
      \centering
      \includegraphics[width=7.5cm,clip]{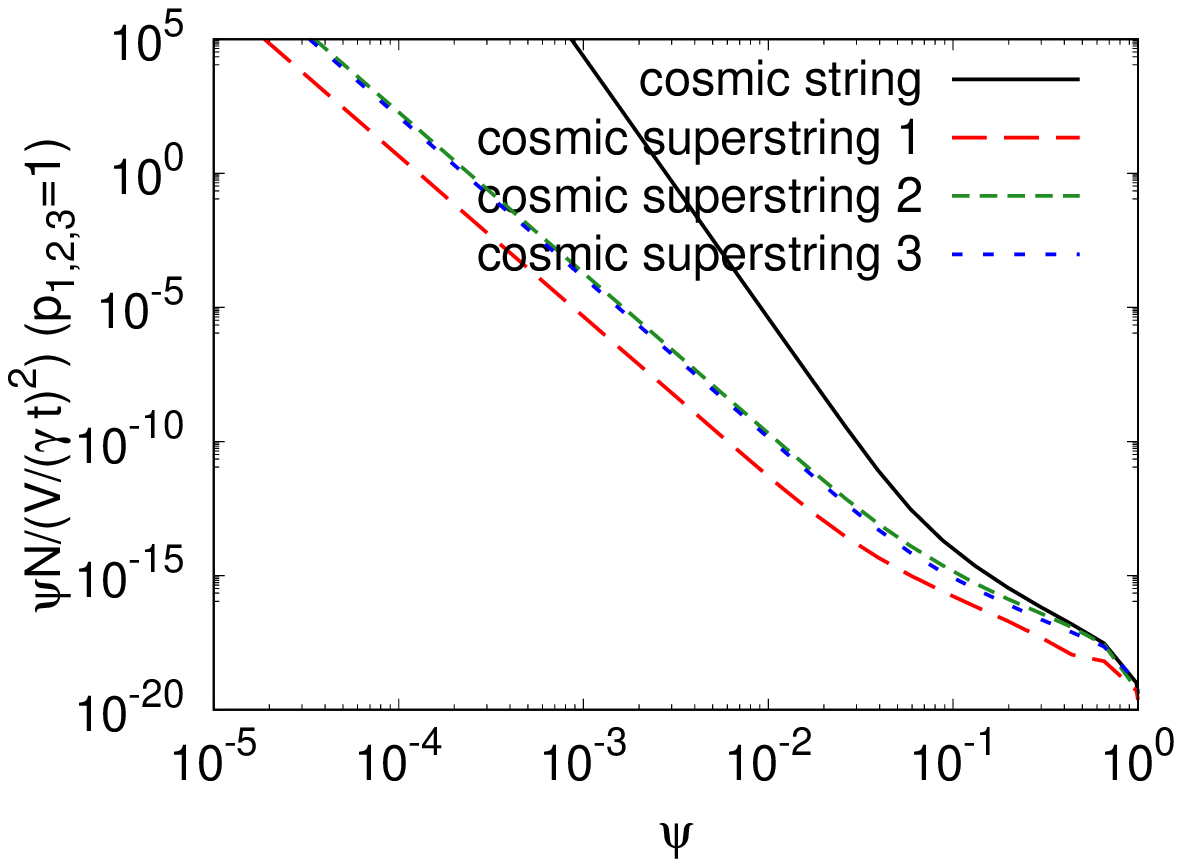}
    \end{minipage}
    \begin{minipage}{0.5\hsize}
      \centering
      \includegraphics[width=7.5cm,clip]{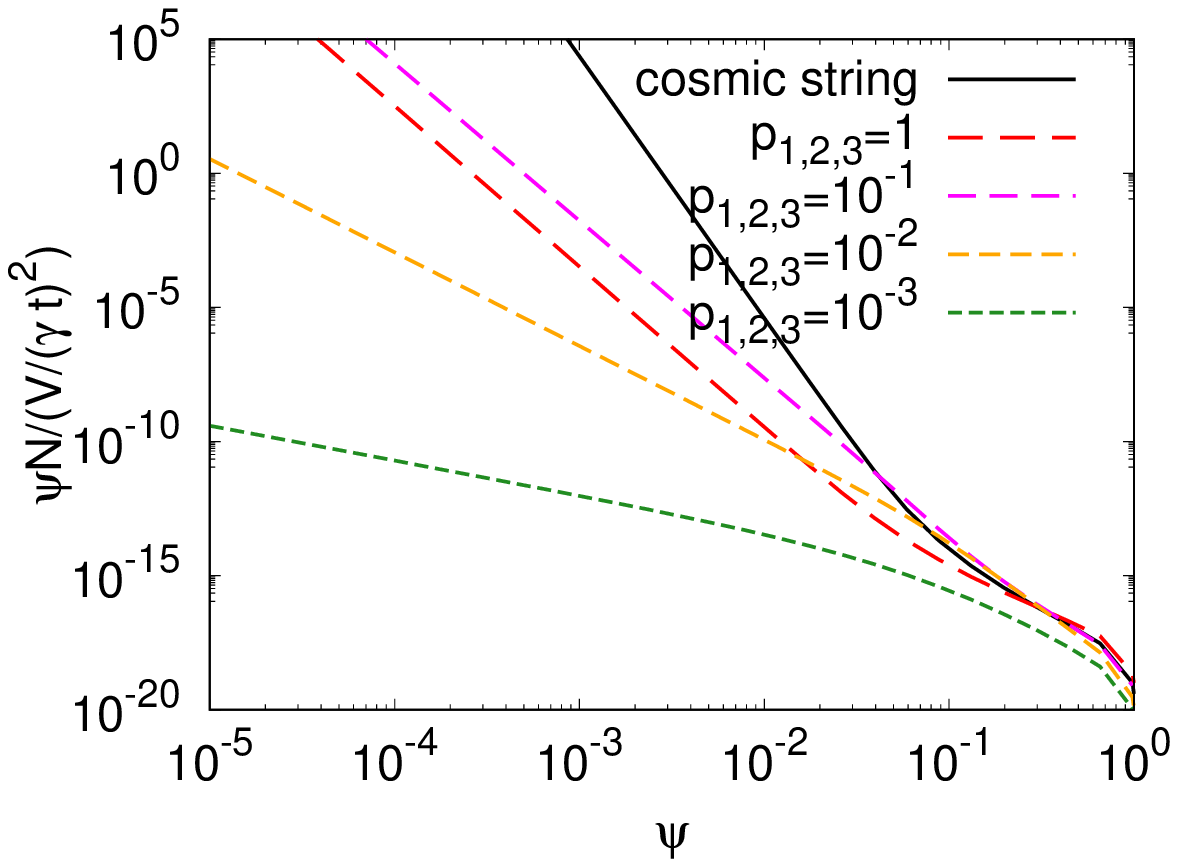}
    \end{minipage}
  \end{tabular}
  \caption{The distribution function of kinks on infinite cosmic superstrings. The axes are same as in Fig.~\ref{fig:kink_distribution_CS}.
    In the left panels, we show results for different string types $1$, $2$ and $3$ (red, green and blue broken lines) for $p_1 = p_2 = p_3 = 1$.
    The right panels show results for different reconnection probabilities.
    The red, magenta, orange and green broken lines correspond to the cases of $p_1=p_2=p_3=1, \, 10^{-1}, \, 10^{-2}$ and $10^{-3}$, respectively. The lines represent the total number of kinks on string types $1$, $2$ and $3$.
   In all panels, for comparison, we plot the case of ordinary cosmic strings with the black solid line.}
\label{fig:kink_distribution_CSS}
\end{figure}

\section{GW background from propagating kinks} \label{sec:propagating_kink}
GWs emitted from kinks overlap one another and form the GW background.
In this section, we calculate the GW background spectrum from kink propagation.
First, we briefly review the calculation method. For details, see \cite{Matsui:2016xnp,Matsui:2019obe}.

\subsection{Formulation}
It has been shown that, for a given frequency of the GW background $f$, kinks whose interval is the same as the wavelength of the GW background contribute most to the GW amplitude \cite{Kawasaki:2010yi}.
Such kinks satisfy the following equation
\begin{equation}
\left (\psi \frac{N(\psi, \, t)}{V(t)/(\gamma t)^2} \right )^{-1} \sim [(1+z)\omega]^{-1} \label{eq:psimax}.
\end{equation}
where $f = \omega/2 \pi$. We denote the sharpness satisfying this equation as $\psi_{\rm m}$, which is a function of frequency and time.

Considering that the GW background of frequency $f$ is formed by kinks with sharpness $\psi_{\rm m}$, the power spectrum of the GW background is written as
\begin{equation}
 \Omega_{\rm GW}(f) \equiv \frac{1}{\rho_{\rm c}} \frac{{\rm d} \rho_{\rm GW}}{{\rm d} \, {\rm ln} f} 
= \frac{2\pi^2f^2}{3H_0^2} \int \frac{dz}{z}\Theta(n(\psi_{\rm m},f,z)-1) n(\psi_{\rm m},f,z)h^2(\psi_{\rm m},f,z), \label{eq:Omega_gw}
\end{equation}
where $\rho_{\rm c}\equiv \frac{3H_0^2}{8\pi G}$ is the critical density, $\rho_{\rm gw}$ is the energy density of the GW background, $\Theta$ is a step function introduced to remove rare bursts.

The strain amplitude of a GW burst from a propagating kink is given by \cite{Damour:2001bk}
\begin{equation}
  h(\psi,f,z)=\frac{\psi^{1/2} G\mu  \gamma t}{\{(1+z)f \gamma t \}^{2/3}}\frac{1}{r(z)} \Theta(1-\theta_m), \label{eq:many_kinks_h_eq}
\end{equation}
where the distance from the observer is given by $r(z) = \int_{0}^{z} \frac{{\rm d} z'}{H(z')}$, and $\theta_m \equiv \{(1+z)f \gamma t \}^{-1/3}$. The step function $\Theta(1-\theta_m)$ is introduced to exclude GWs whose wavelength is larger than the curvature radius of a string.
Note that the GW amplitude depends on the sharpness of the kink as $\propto \psi^{1/2}$, and thus kinks with large sharpness produce stronger GWs.

The effective GW rate $n(\psi_{\rm m},f,z)\equiv \frac{1}{f}\frac{{\rm d}\dot{N}}{{\rm d}\ln z}$, where $\dot{N}$ is the event rate of GW burst with frequency $f$ at redshift $z$, is given by \cite{Matsui:2016xnp}
\begin{equation}
  n(\psi_{\rm m},f,z) = \frac{1}{f}\frac{\theta_m}{2(1+z)} \frac{1}{\gamma t}\frac{\psi_{\rm m}N(\psi_{\rm m},t)}{V}\frac{{\rm d} V(z)}{{\rm d} \ln z}.
\label{eq:n_eq}
\end{equation} 
The volume per redshift interval is $\frac{{\rm d} V}{{\rm d} z} = \frac{1}{z}\frac{{\rm d} V}{{\rm d} \ln z} = \frac{4\pi a^3r^2(z)}{H(z)}$.
By combining Eqs.~\eqref{eq:n_eq} and \eqref{eq:psimax}, we can find $ \Omega_{\rm GW}(f)$ has the dependence of $\gamma^{-8/3}$. Therefore, the power spectrum becomes stronger with shorter correlation length.

\subsection{Results}
Using the distribution function of kinks obtained in Sec. \ref{sec:kink_distribution}, we numerically calculate the power spectrum of the GW background $\Omega_{\rm GW}$, and the results are shown in Fig. \ref{fig:kink_GWB_CSS}. The left panels of Fig.~\ref{fig:kink_GWB_CSS} are calculated assuming $p_1 = p_2 = p_3 = 1$ for different string types $1$, $2$, and $3$. 
The right panels show the results for different reconnection probabilities. From the top to the bottom, we show Case A, B, and C.

\subsubsection{Case A: GW background with $\mu_1:\mu_2:\mu_3=1:1:1$ and $n_p=1$}
\label{subsec:GWB_mu:mu:mu_np=1}
Let us first see the left panel, which is helpful to see the pure effect of Y-junctions. We see the power spectrum of ordinary cosmic strings, which is shown for comparison, increases gradually towards high frequencies. 
The GW background is mainly formed by GWs emitted from kinks existing today and today's kink distribution determines the spectral shape. For detailed explanations, see \cite{Matsui:2016xnp}. 
On the other hand, in Case A, we see the spectra of cosmic superstrings are flat at high frequencies and the amplitude is lower than the cosmic string case. The difference in shape arises because, in the case of cosmic superstrings, the dominant contribution to the GW amplitude comes from GWs emitted from kinks in the old time just after their formation when $\psi\sim 1$. As we have seen in the previous section, sharp kinks are smoothed out rapidly because of Y-junctions, and since the GW strain amplitude depends on $\propto \psi^{1/2}$, kinks today with very small sharpness no longer contribute to the GW background.

Since kinks are formed by collisions of infinite strings, when a new kink with $\psi\sim 1$ is formed, the GWs emitted soon after the kink formation have a wavelength of order of the curvature radius of strings $\sim \gamma t$ and thus we have the relation of $(1+z)f\gamma t \sim 1$. From this, we see that higher frequency GWs are produced by kinks in earlier times of the Universe. Since the number of infinite strings in the Hubble horizon is always the same because of the scaling law, the number of newly formed kinks inside the horizon is also the same. This means that, if we only consider GWs emitted from new kinks, the energy ratio $\rho_{\rm GW}/\rho_c$ is constant in time, which is the reason for the flat shape of the spectrum.  

The flat spectrum can be also explained by using equations. Let us assume that the steepness of kink distribution as a function of sharpness is given by the power-law as $\propto \psi^{-l}$. In \cite{Kawasaki:2010yi}, it has been shown that the kink number decreases in time as $\propto t^{-1}$ in the RD era. Thus, we can write the number distribution in the RD era as 
$\psi \frac{N(\psi, \, t)}{V(t)/(\gamma t)^2} \propto \psi^{-l} t^{-1}$.
Substituting this into Eq.~\eqref{eq:psimax}, we find $\psi_{\rm m}$ can be described as
\begin{equation}
  \psi_{\rm m} \propto [(1+z)ft]^{-1/l}.
  \label{eq:psimax2}
\end{equation} 
Considering that the GW mode contributing to the background amplitude satisfies $(1+z)f\gamma t \sim 1$ and $\gamma$ is constant because of the scaling law, we find $\psi_{\rm m}$ is determined independently of the frequency and time.
Using Eq.~\eqref{eq:psimax} and $\psi_{\rm m}(f,t) = {\rm const.}$, we find $\Omega_{\rm GW} \propto f^0$. The flat spectrum is produced by GWs from the RD era, while the increase of GW amplitude at low frequencies corresponds to GWs generated in the MD era.

The right panel of the figure shows the cases of different reconnection probabilities. We find that the power spectrum becomes smaller for a smaller reconnection probability. This can be explained by the balance between the correlation length and $\psi_{\rm m}$. As we can see from Eq.~\eqref{eq:psimax2}, when the slope of the kink distribution is flattened and $l$ is small, the value of $\psi_{\rm m}$ becomes very small. This means that kinks contributing to the GW background have very small sharpness, and since $\Omega_{\rm GW}$ has the dependence of $h^2 \propto \psi_{\rm m}$, the amplitude of GWs drops. As shown in Fig.~\ref{fig:kink_distribution_CSS}, the slope of the distribution function get gentler for a smaller reconnection probability in Case A, which leads to a smaller GW amplitude. At the same time, the power spectrum has the dependence of $\gamma^{-8/3}$ and a smaller reconnection probability decreases the value of $\gamma$ and increases the GW amplitude, but the effect of $\psi_{\rm m}$ dominates in Case A. 

In the figure, we find that the low-frequency cutoff moves towards high frequency. This is because of the cutoff $\Theta(1-\theta_m)$, which prohibits the emission of GWs with a wavelength longer than the curvature radius of strings. A small reconnection probability makes the correlation length short and the curvature of strings gets smaller so that we do not find GWs at low frequency.

\subsubsection{Case B: GW background with $\mu_1:\mu_2:\mu_3=1:1:1$ and $n_p=\frac{1}{3}$}
\label{subsec:GWB_mu:mu:mu_np=1over3}
In Case B, the power spectrum looks similar to the one of Case A and the reasons are the same as explained for Case A. In the right panel, we find that the GW amplitude decreases more for a smaller reconnection probability compared to Case A. This is because, the loop production is more efficient in the case of $n_p=\frac{1}{3}$, and the decrease of correlation length $\gamma$ is milder compared to Case A as shown in the right panels of Fig.~\ref{fig:scaling_CSS_different_p}. Since the value of $\gamma$ does not decrease, a more prominent effect of $\psi_{\rm m}$ is seen, which turns into a smaller amplitude of the GW background.

\subsubsection{Case C: GW background with $\mu_1:\mu_2:\mu_3=1:10:10$ and $n_p=\frac{1}{3}$}
\label{subsec:GWB_mu:10mu:10mu_np=1over3}
In the left panel of Case C, the shape of the power spectrum looks similar to the ordinary cosmic strings. This is because the effect of Y-junctions to smooth out kink sharpness is gentler in Case C, and the dominant contribution to the GW power is made by kinks existing today.
We find that strings $2$ and $3$ generate the larger GW amplitude than string $1$, since they have larger string tension and the power spectrum has the dependence of $\propto(G\mu)^2$. We also find string $3$ has a slightly larger amplitude compared to string $2$, because the correlation length of string $3$ is smaller than the others and the power spectrum has the dependence of $\propto \gamma^{-8/3}$.

In the right panel, we find the interesting cases where the power spectrum is slightly larger than the ordinary cosmic string case. This is because the slope of the kink distribution is not entirely flattened compared to Cases A and B as shown in Fig.~\ref{fig:kink_distribution_CSS} and the value of $\psi_{\rm m}$ is relatively large. We find that the increase of the power spectrum with the dependence of $\gamma^{-8/3}$ dominates the effect of small $\psi_{\rm m}$ in the case of $p=10^{-1}$ and $10^{-2}$, while the kink distribution becomes too gentle when $p=10^{-3}$ and the latter effect dominates the former.  

\begin{figure}[htbp]
  \centering
  \begin{tabular}{c}
    Case A: $\mu_1:\mu_2:\mu_3=1:1:1$, $n_p=1$, $G\mu_1=G\mu_2=G\mu_3=10^{-11}$ \\
    \begin{minipage}{0.5\hsize}
      \centering
      \includegraphics[width=7.5cm,clip]{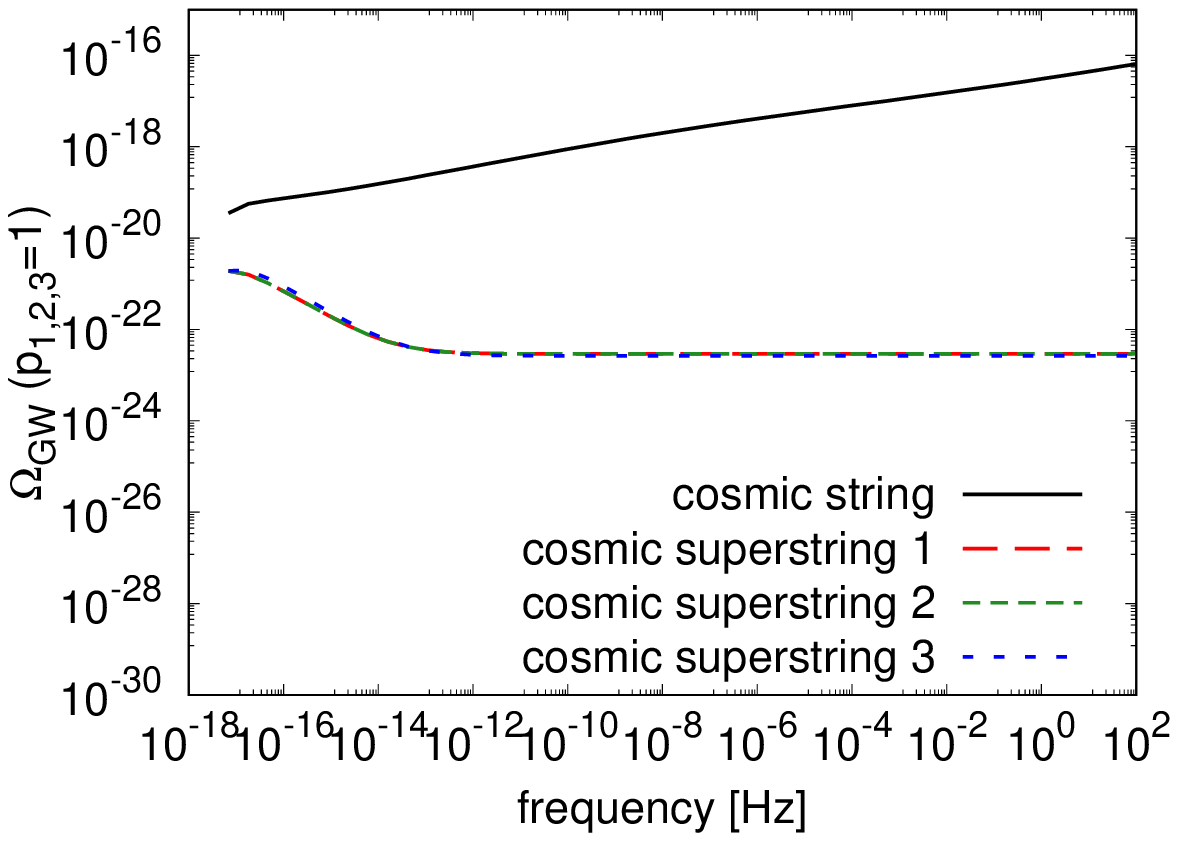}
      \end{minipage}  
    \begin{minipage}{0.5\hsize}
      \centering
      \includegraphics[width=7.5cm,clip]{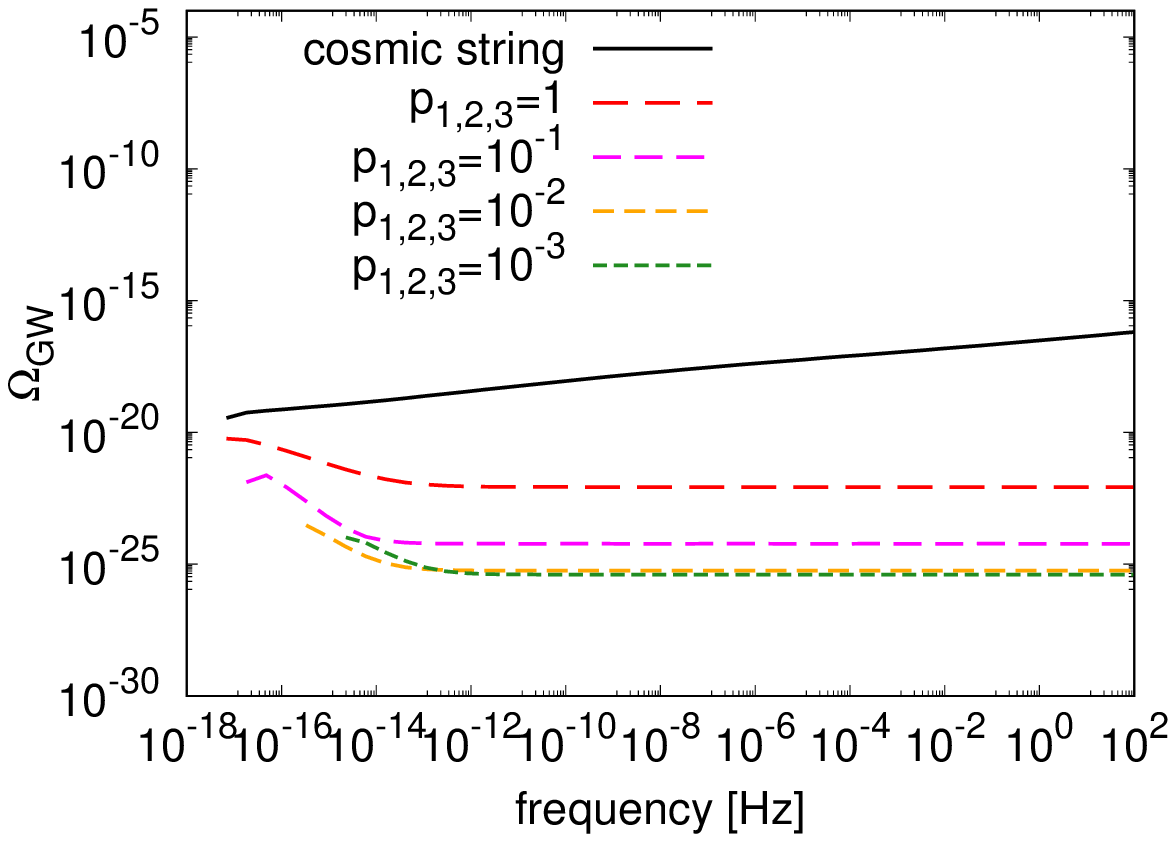}
    \end{minipage} \vspace{20pt} \\
    Case B: $\mu_1:\mu_2:\mu_3=1:1:1$, $n_p=\frac{1}{3}$, $G\mu_1=G\mu_2=G\mu_3=10^{-11}$ \\
        \begin{minipage}{0.5\hsize}
      \centering
      \includegraphics[width=7.5cm,clip]{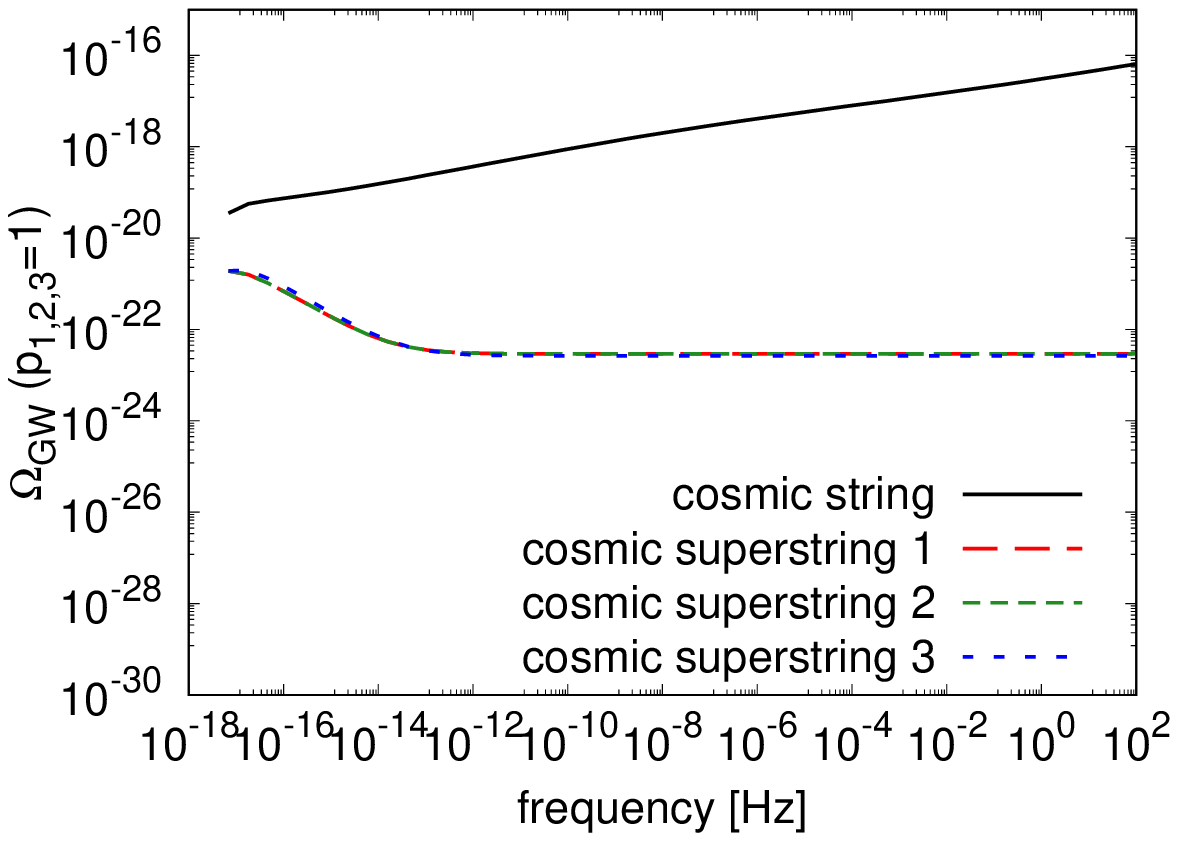}
    \end{minipage} 
    \begin{minipage}{0.5\hsize}
      \centering
      \includegraphics[width=7.5cm,clip]{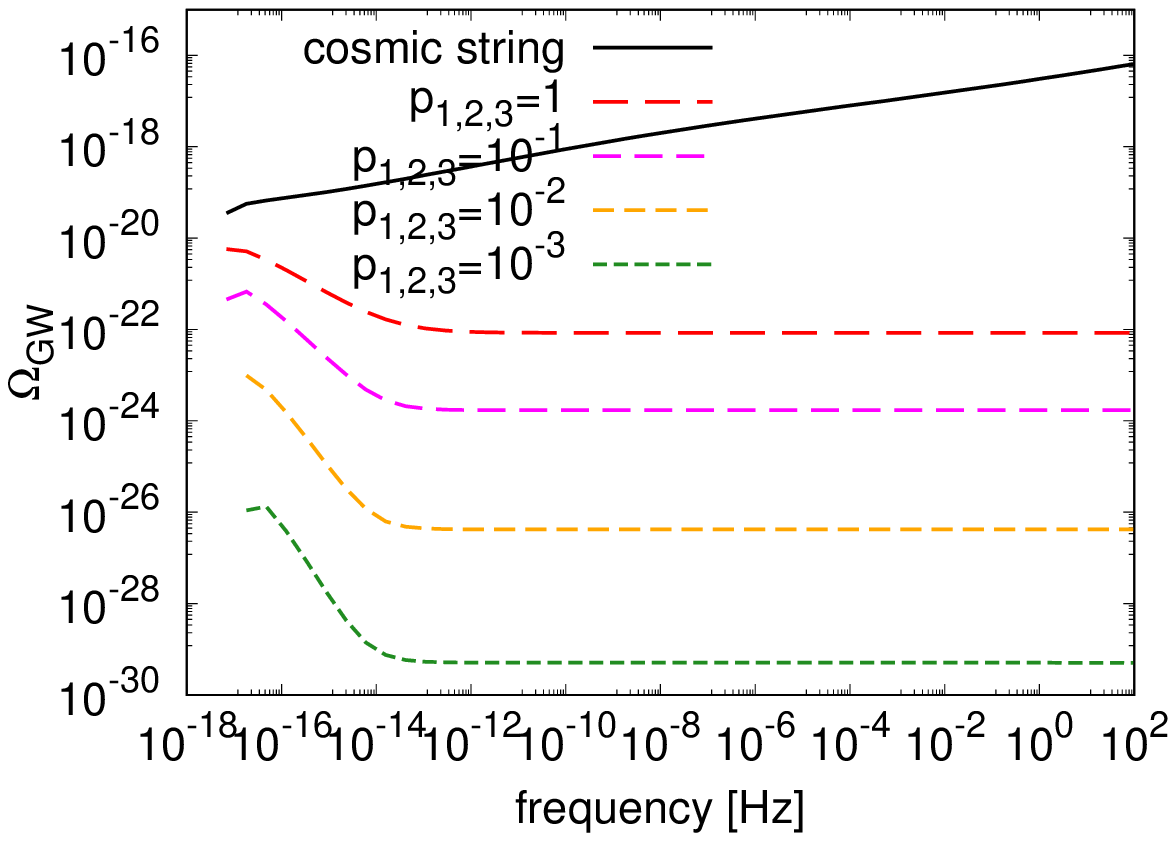}
    \end{minipage} \vspace{20pt} \\
    Case C: $\mu_1:\mu_2:\mu_3=1:10:10$, $n_p=\frac{1}{3}$, $G \mu_1 = 10^{-12}, \, G \mu_2 = G \mu_3 = 10^{-11}$ \\
    \begin{minipage}{0.5\hsize}
      \centering
      \includegraphics[width=7.5cm,clip]{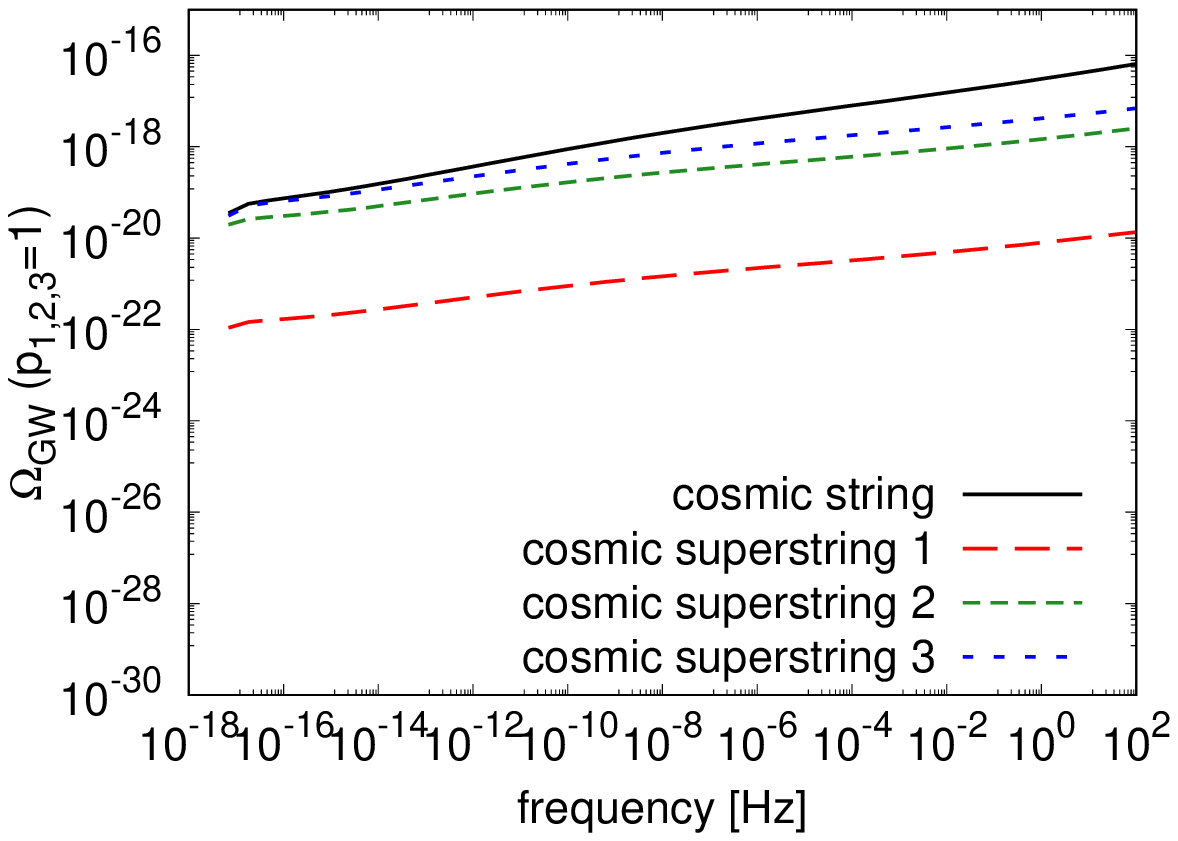}
    \end{minipage}
    \begin{minipage}{0.47\hsize}
      \centering
      \includegraphics[width=7.5cm,clip]{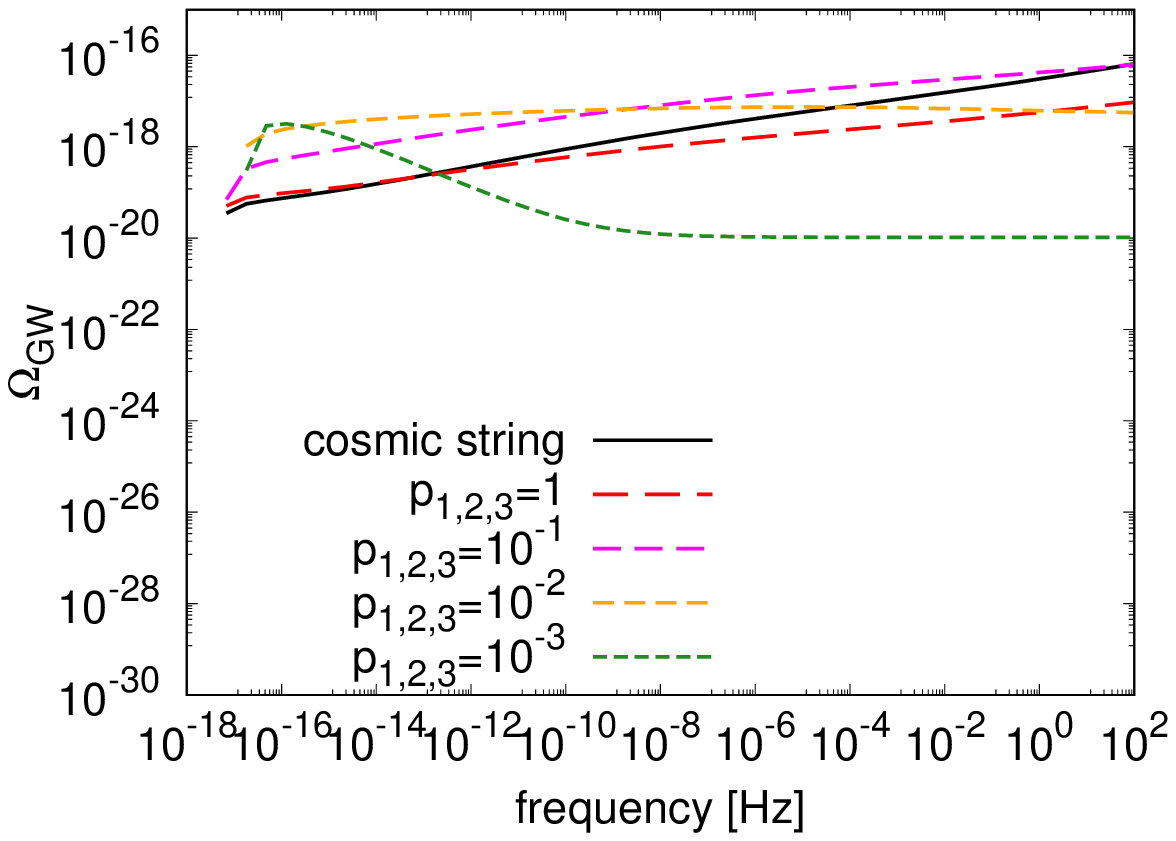}
    \end{minipage}
  \end{tabular}
  \caption{The power spectrum of the GW background from propagating kinks on infinite cosmic superstrings. The vertical axis is the spectral amplitude of the GW background $\Omega_{\rm GW}$ and the horizontal axis is the frequency. Left panels show the case of $p_1=p_2=p_3=1$ for different string types. 
The red, green and blue broken lines represent string types $1$, $2$, and $3$, respectively.
The right panels are the power spectrum of the GW background for different reconnection probabilities.
The red, magenta, orange and green broken lines correspond to $p=1, \, 10^{-1}, \, 10^{-2}$ and $10^{-3}$, respectively. In all panels, for comparison, we plot the case of ordinary cosmic strings with the black solid line ($G\mu = 10^{-11}$).}
\label{fig:kink_GWB_CSS}
\end{figure}

\section{GW background from kink-kink collisions} \label{sec:kink-kink_collision}
In this section, we calculate the power spectrum of the GW background from kink-kink collisions. First we briefly review the calculation method. For details, see \cite{Matsui:2019obe}. 
\subsection{Formulation}
The GW spectrum for kink-kink collisions can be calculated by replacing $n(\psi_{\rm m},f, z)$ and $h^2(\psi_{\rm m},f,z)$ in Eq. \eqref{eq:Omega_gw}. In the case of kink-kink collisions, the strain amplitude is given by \cite{Binetruy:2009vt}
\begin{equation}
h_{\rm kk}(\psi,f,z) = \frac{\psi G \mu}{(1+z) f} \frac{1}{r(z)} \Theta(1-\theta_m). 
\label{eq:k-k_col_h_eq}
\end{equation} 
The effective GW rate is given by \cite{Matsui:2019obe}
\begin{equation}
n_{\rm kk}(\psi_{\rm m},f,z) = \frac{1}{f} \frac{(\gamma t)^2}{2(1+z)} \Bigl\{\frac{\psi_{\rm m} N(\psi_{\rm m},t)}{V} \Bigr\}^2  \frac{{\rm d}V(z)}{{\rm d} \, {\rm ln} z}. 
\label{eq:nkk}
\end{equation} 
Note that the event rate is boosted by the square of the kink number. Since kinks with small sharpness are numerous and they produce high-frequency GWs, we can expect a large amplitude of the GW background at high frequency.
In fact, in our previous work \cite{Matsui:2019obe}, we found that the GW amplitude is dramatically enhanced at high frequencies and the substantial amount of GW emissions could affect the scaling behavior of the string network as well as the shape of the kink distribution as backreaction effects. By following the method established in \cite{Matsui:2019obe}, we take into account the backreaction effects of the large GW emission as follows.

First, we modify the evolution equation of the correlation length, Eq.~\eqref{eq:L_eq}, to include the energy loss of the string network through GW emissions as 
\begin{equation}
    \frac{{\rm d}L}{{\rm d}t} = HL(1 +v^2) +\frac{1}{2}cp^{n_p}v +\frac{\pi^3 G \mu}{2} \gamma t \int_{0}^{1} {\rm d}\psi_{\rm m} \, \frac{N(\psi_{\rm m},t)}{V/(\gamma t)^2} \psi_{\rm m}^2 . 
    \label{eq:L_eq_with_GWR}
\end{equation}
The last term of the right hand side describes the energy loss of the string network through GW emissions.
In the same manner, the GW term is added in the equations for cosmic superstrings, Eqs. \eqref{eq:L_1_CSS_eq}, \eqref{eq:L_2_CSS_eq}, and \eqref{eq:L_3_CSS_eq}.

Second, the backreaction of GW emissions could smooth out the sharpness of kinks. We modify the evolution equation of the kink distribution, Eq. \eqref{eq:kink_eq}, as 
\begin{equation}
  \frac{\partial N}{\partial t}(\psi, \, t) = \frac{\bar{\Delta} V}{\gamma^4 t^4} g(\psi) +\frac{2 \zeta}{t} \frac{\partial}{\partial \psi} (\psi N(\psi, \, t)) -\frac{\eta}{\gamma t} N(\psi, \, t) -\frac{\pi^3 G \mu \psi^2(\gamma t)^2}{V} N^2(\psi, \, t). 
  \label{eq:kink_CS_eq_with_BR}
\end{equation} 
The last term gives the number of kinks lost by GW emissions. This term is also added for the superstring case in Eqs. \eqref{eq:kink_eq_CSS_1}, \eqref{eq:kink_eq_CSS_2}, and \eqref{eq:kink_eq_CSS_3}. 
 
\subsection{Results}
We numerically calculate the power spectrum using Eqs.~\eqref{eq:k-k_col_h_eq} and \eqref{eq:nkk} by taking into account the backreaction effects of the GW emission using Eqs.~\eqref{eq:L_eq_with_GWR} and \eqref{eq:kink_CS_eq_with_BR}. In Fig. \ref{fig:kink-kink_collison_distribution_and_GWB_CSS_BR}, we show the kink distribution on the left panels. The settings are the same as the right panels of Fig.~\ref{fig:kink_distribution_CSS} except here we include the backreaction effects of the GW emission by kink-kink collisions.
The right panels show the power spectra of the GW background from kink-kink collisions.

\subsubsection{Case A: GW background with $\mu_1:\mu_2:\mu_3=1:1:1$ and $n_p=1$}
Let us first see the kink distribution function, shown in the left panel. Compared to Fig.~\ref{fig:kink_distribution_CSS}, we find that the number of kinks with small sharpness is suppressed only in the case of ordinary cosmic strings. This occurs because kinks with small sharpness are numerous and a large number of their collisions turn on the GW backreaction effect through Eq.~\eqref{eq:kink_CS_eq_with_BR}. The correlation length also becomes large because of \eqref{eq:L_eq_with_GWR}, and the slope of the distribution function gets slightly gentler. On the other hand, in the case of cosmic superstrings, the number of kinks is reduced by Y-junctions and the GW amplitude is suppressed so that the backreaction effect of the GW emission is too small.

In the right panel, in the case of cosmic strings, we find that the GW spectrum increases towards high frequency and the spectrum becomes flat at around $10^{-7}$Hz. The flat behavior at a high frequency is because of the GW backreaction. See \cite{Matsui:2019obe}, for details. In the case of cosmic superstrings, we find the spectral amplitude is low and has a flat spectrum, since Y-junctions smooth out kink sharpness and the value of $\psi_{\rm m}$ becomes very small. We also find that the overall GW power decreases for a smaller reconnection probability. The reason is similar to the case of propagating kinks, explained in Sec. \ref{subsec:GWB_mu:mu:mu_np=1}.
In the case of kink-kink collisions, the spectral amplitude depends on the correlation length and $\psi_{\rm m}$ as $\Omega_{\rm GW} \propto \gamma^{-2} \psi_{\rm m}^2$.
In Case A, the effect of $\psi_{\rm m}$ dominates the one of $\gamma$ for a small reconnection probability.

\subsubsection{Case B: GW background with $\mu_1:\mu_2:\mu_3=1:1:1$ and $n_p=\frac{1}{3}$}
The results of Case B are similar to Case A. The difference appears when the reconnection probability is small, where we find the GW amplitude decreases more. As explained in Sec.~\ref{subsec:GWB_mu:mu:mu_np=1over3}, this is because the decrease of correlation length $\gamma$ is milder compared to Case A and a more prominent effect of $\psi_{\rm m}$ is seen.

\subsubsection{Case C: GW background with $\mu_1:\mu_2:\mu_3=1:10:10$ and $n_p=\frac{1}{3}$}
In the same way as GWs from kink propagation in Sec.~\ref{subsec:GWB_mu:10mu:10mu_np=1over3}, we find that the power spectrum is slightly enhanced compared to the ordinary cosmic string case when $p=10^{-1}$. This is again because the slope of the kink distribution is not entirely flattened by Y-junctions compared to Cases A and B and the value of $\psi_{\rm m}$ is relatively large. The increase of the GW amplitude by a small correlation length with $\propto \gamma^{-2}$ dominates the effect of small $\psi_{\rm m}$ in the case of $p=10^{-1}$ and $10^{-2}$, while the kink distribution becomes too gentle and the GW amplitude becomes very low when $p=10^{-3}$.

\begin{figure}[htbp]
  \centering
  \begin{tabular}{c}
  Case A: $\mu_1:\mu_2:\mu_3=1:1:1$, $n_p=1$, $G\mu_1=G\mu_2=G\mu_3=10^{-11}$ \\
    \begin{minipage}{0.5\hsize}
      \centering
      \includegraphics[width=7.5cm,clip]{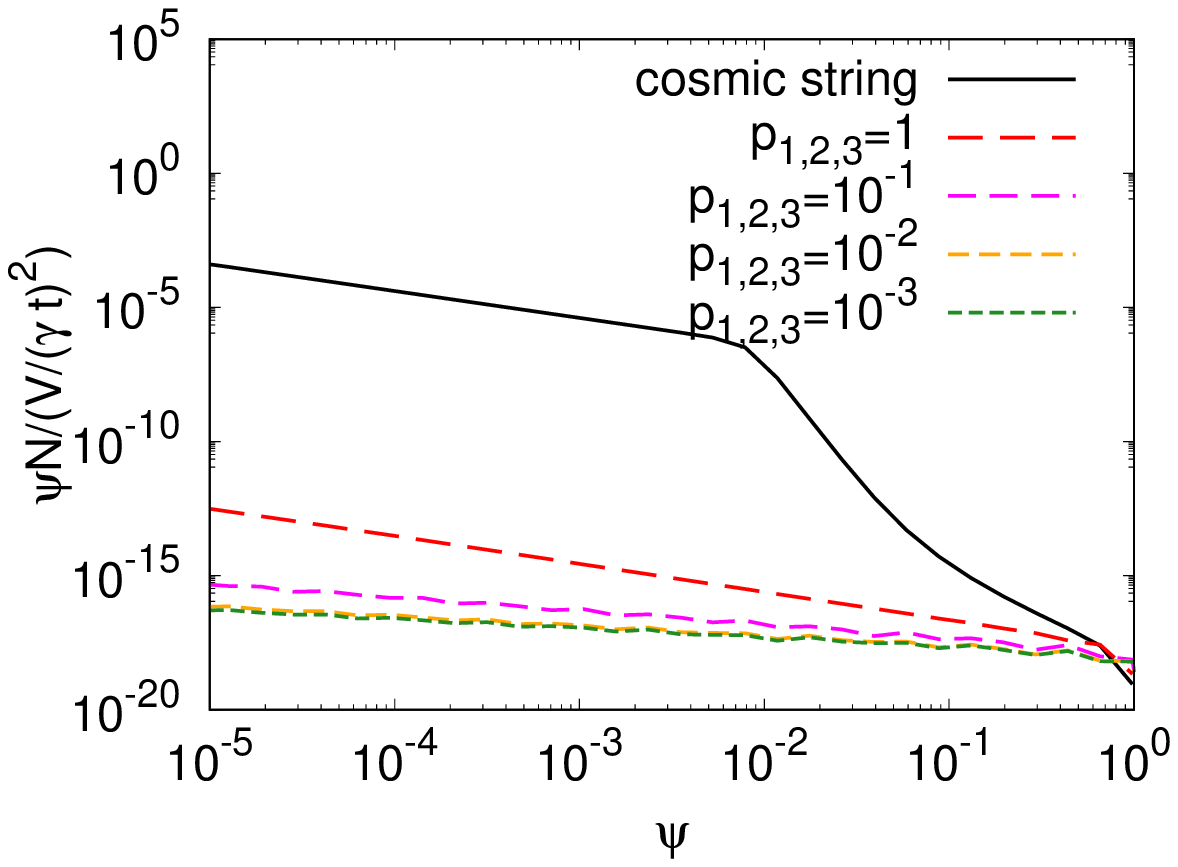}
      \end{minipage}  
    \begin{minipage}{0.5\hsize}
      \centering
      \includegraphics[width=7.5cm,clip]{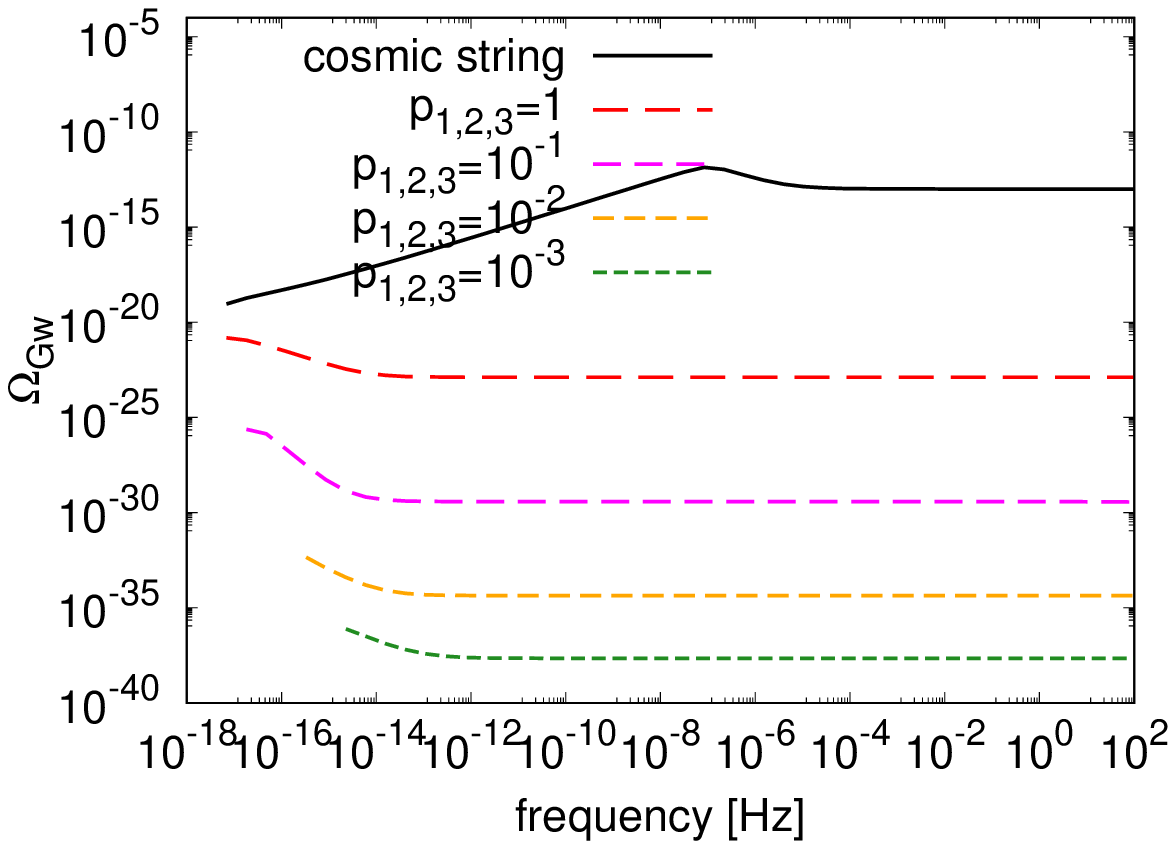}
    \end{minipage} \vspace{20pt} \\
    Case B: $\mu_1:\mu_2:\mu_3=1:1:1$, $n_p=\frac{1}{3}$, $G\mu_1=G\mu_2=G\mu_3=10^{-11}$ \\
        \begin{minipage}{0.5\hsize}
      \centering
      \includegraphics[width=7.5cm,clip]{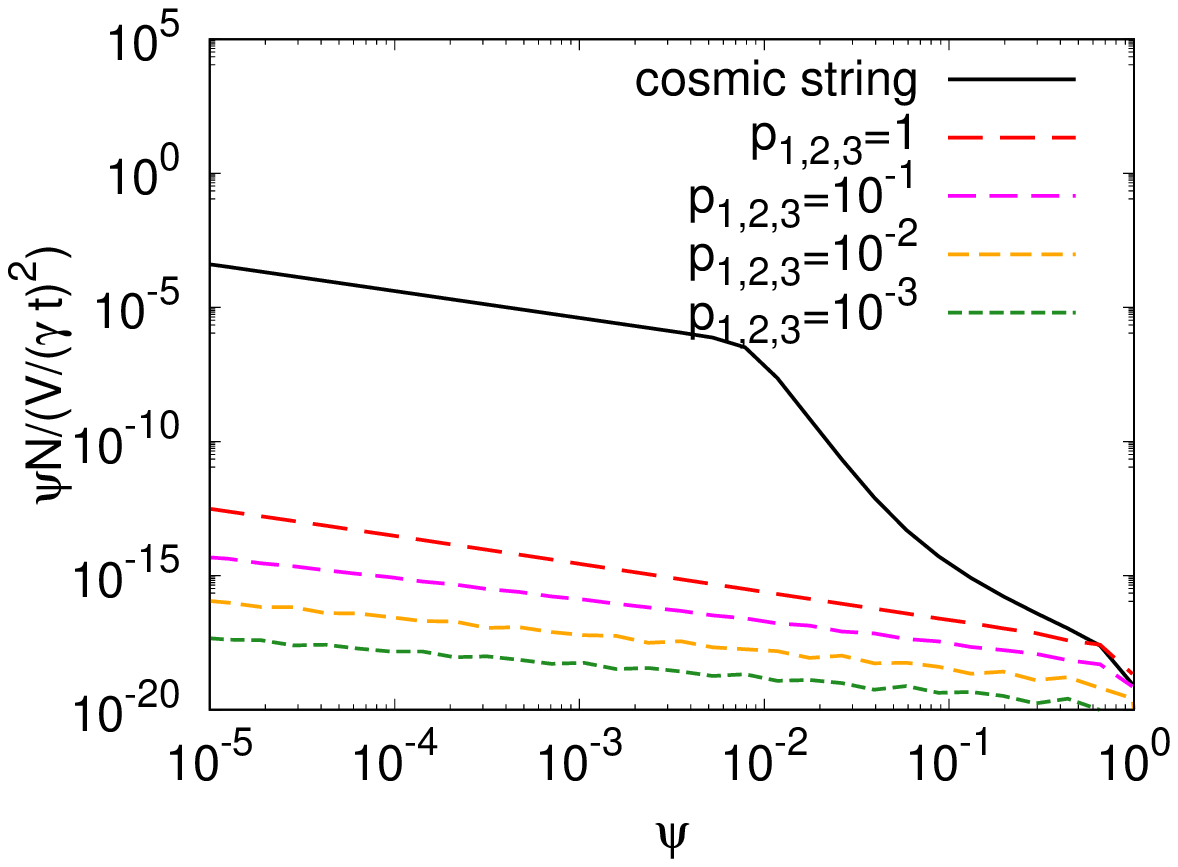}
    \end{minipage}
    \begin{minipage}{0.5\hsize}
      \centering
      \includegraphics[width=7.5cm,clip]{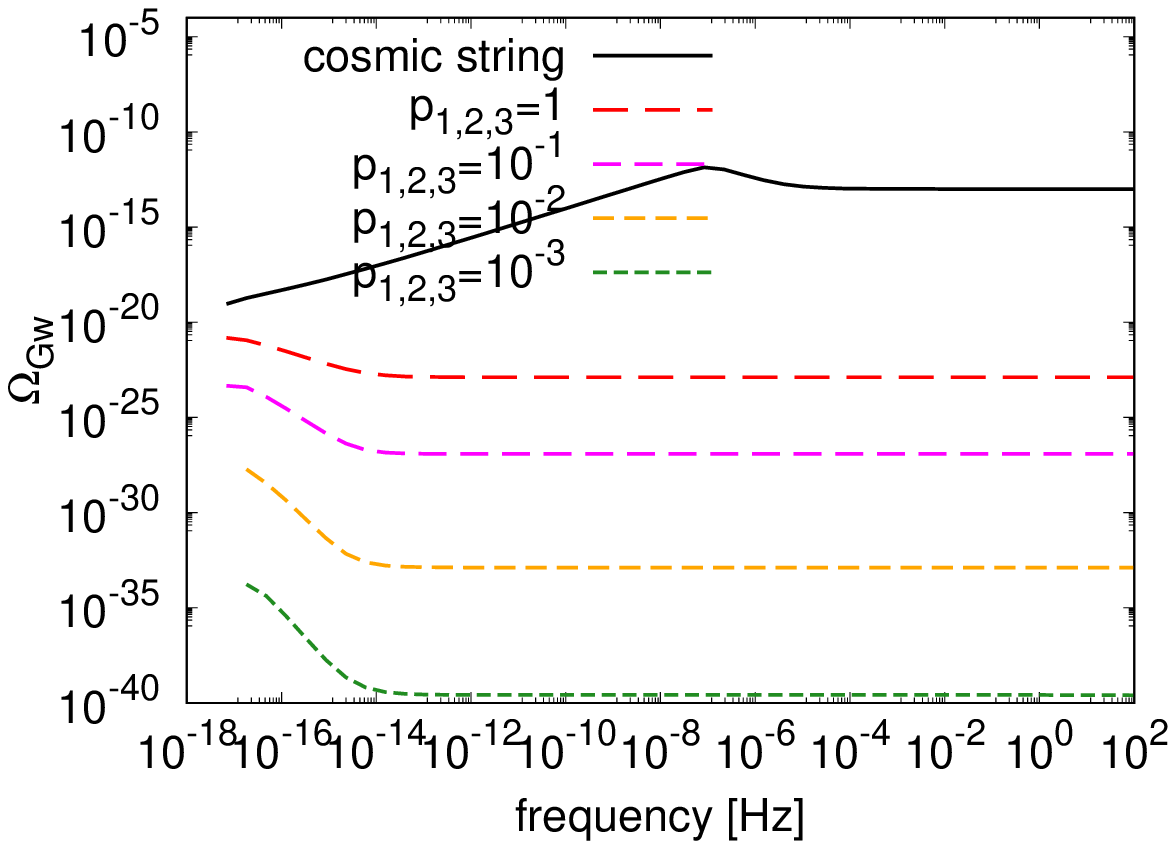}
    \end{minipage} \vspace{20pt} \\
    Case C: $\mu_1:\mu_2:\mu_3=1:10:10$, $n_p=\frac{1}{3}$, $G \mu_1 = 10^{-12}, \, G \mu_2 = G \mu_3 = 10^{-11}$ \\
    \begin{minipage}{0.5\hsize}
      \centering
      \includegraphics[width=7.5cm,clip]{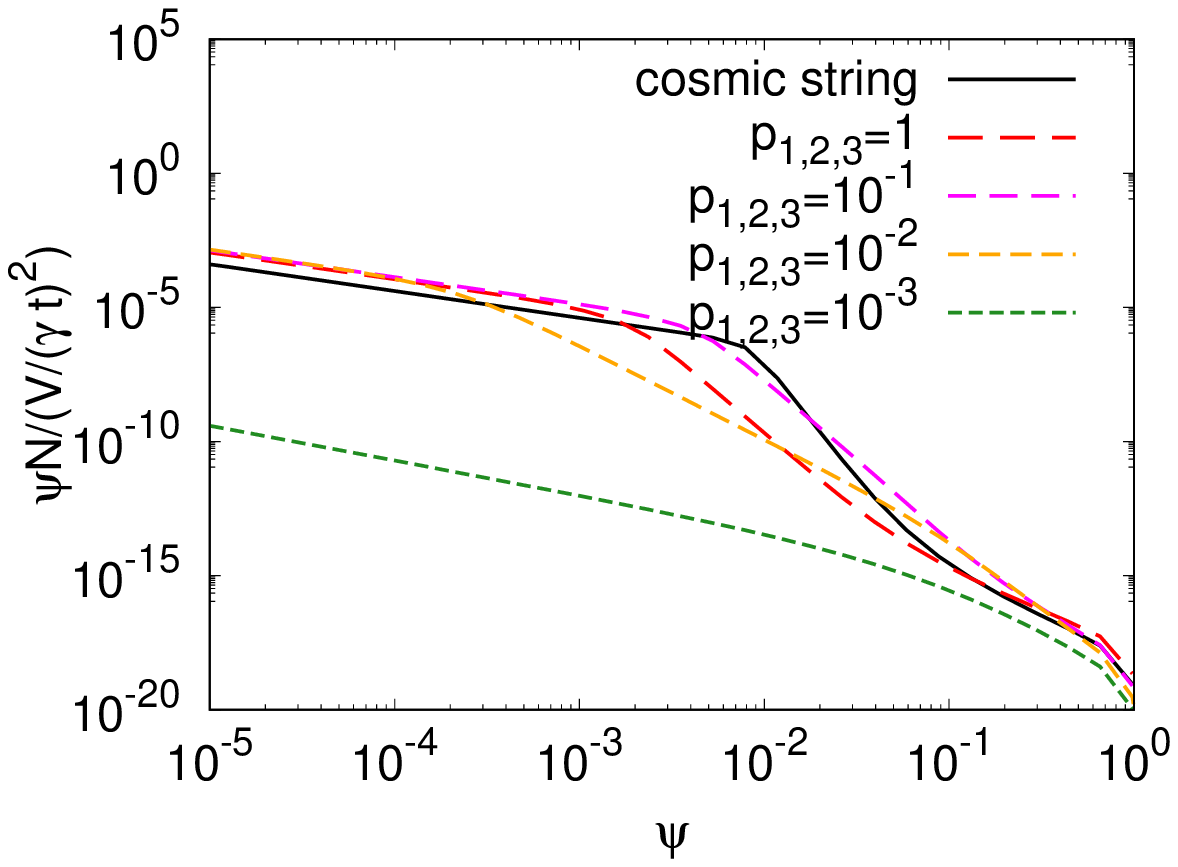}
    \end{minipage}
    \begin{minipage}{0.5\hsize}
      \centering
      \includegraphics[width=7.5cm,clip]{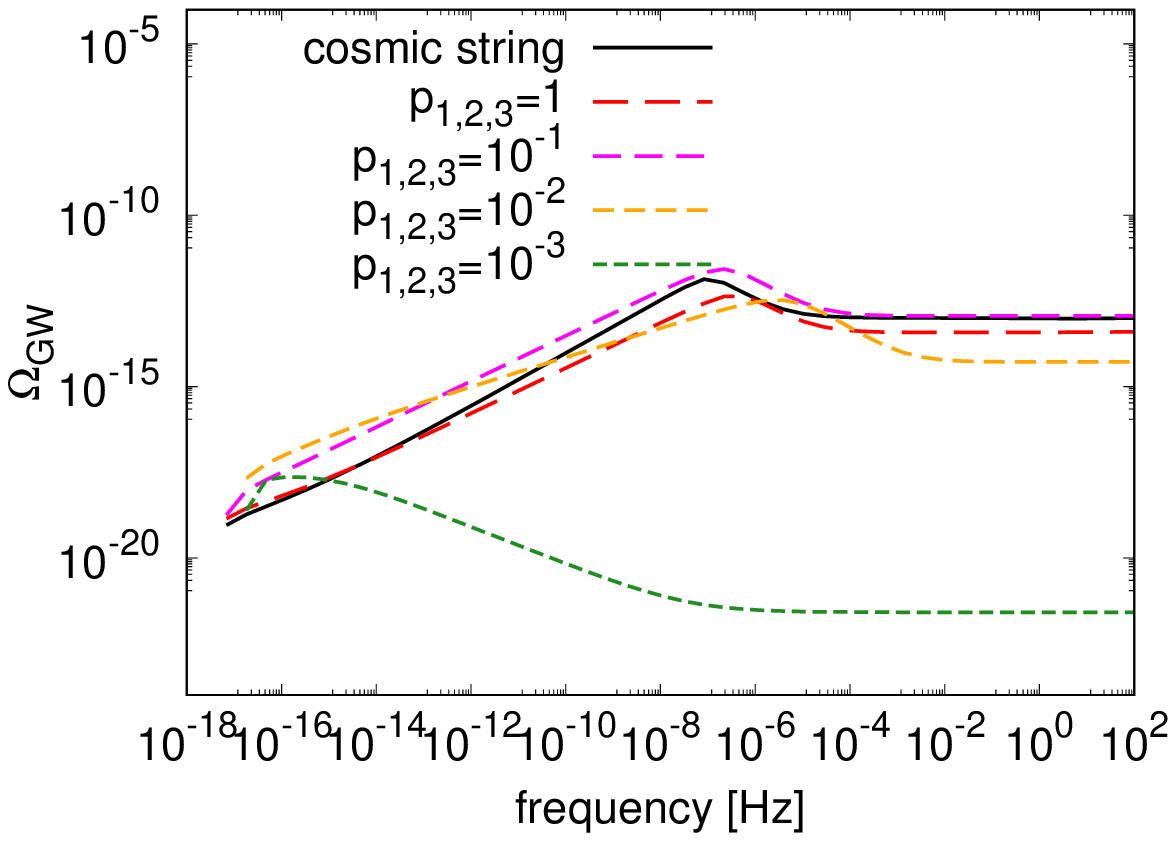}
    \end{minipage}  \vspace{20pt}
  \end{tabular}
  \caption{Left: the distribution function of kinks on infinite cosmic superstrings calculated by taking into account the effects of GW emission. 
  Right: the power spectrum of the GW background from kink-kink collisions on infinite cosmic strings for different reconnection probabilities. The red, magenta, orange and green broken lines represent $p=1, \, 10^{-1}, \, 10^{-2}$ and $10^{-3}$, respectively. In all panels, for comparison, we plot the case of ordinary cosmic strings with the black solid line ($G \mu = 10^{-11}$).}
\label{fig:kink-kink_collison_distribution_and_GWB_CSS_BR}
\end{figure}

\begin{figure}[htbp]
\centering
\includegraphics[width=12cm,clip]{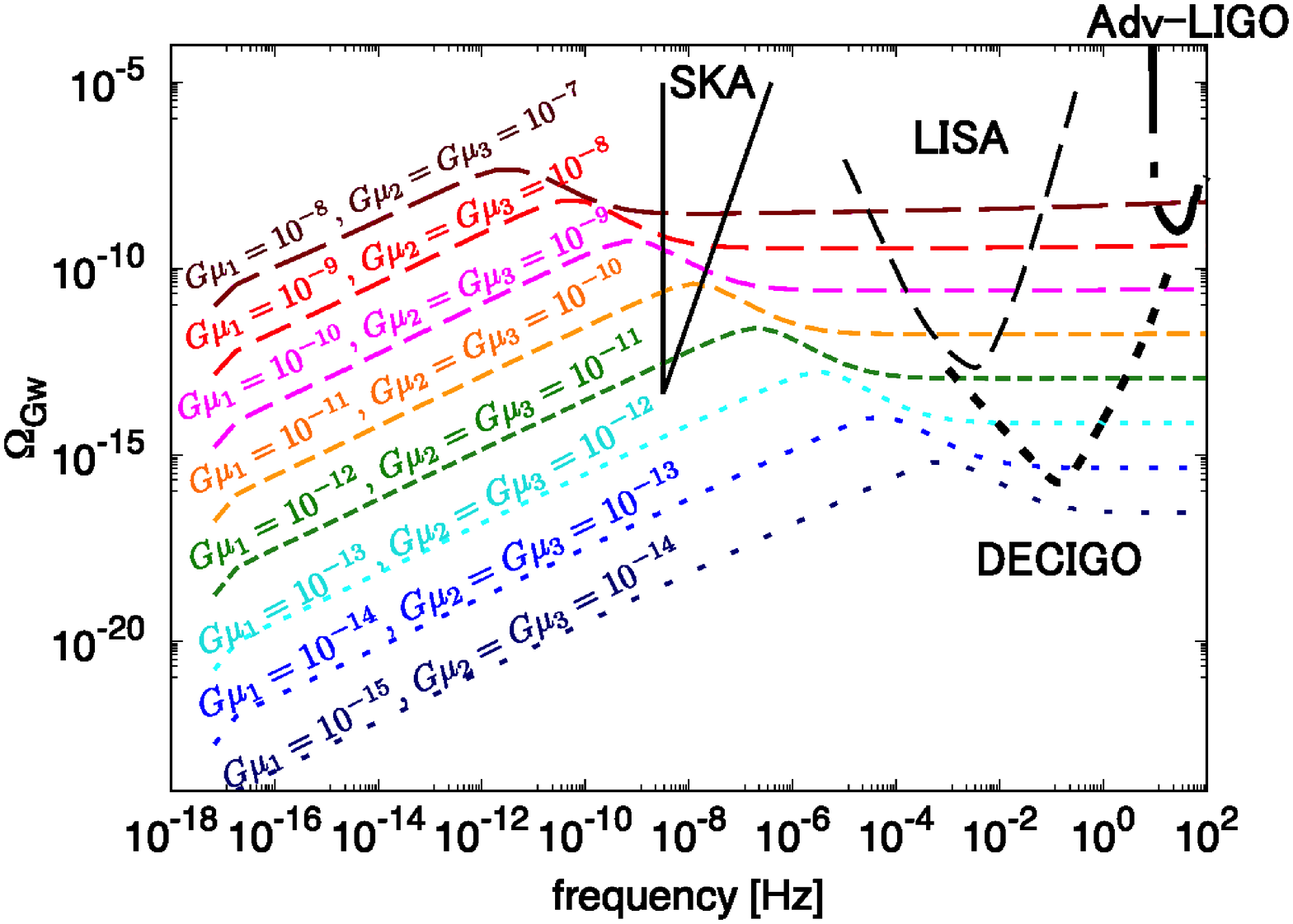}
\caption{The power spectrum of the GW background from kink-kink collisions of Case C ($\mu_1:\mu_2:\mu_3=1:10:10$ and $n_p=\frac{1}{3}$) for different tensions. 
The reconnection probability is set to be $p=10^{-1}$. The black solid and broken lines are the sensitivity curves of future experiments.}
\label{fig:k-k_col_GWB_CSS_BR_different_mu}
\end{figure}

By comparing between Fig.~\ref{fig:kink_GWB_CSS} and \ref{fig:kink-kink_collison_distribution_and_GWB_CSS_BR}, we find that the amplitude of the GW background from kink-kink collisions is larger than the one from kink propagation. If kink sharpness is not smoothed out dramatically by Y-junctions, we may be able to detect GWs from kink-kink collisions by future GW experiments. 
As an example, in Fig. \ref{fig:k-k_col_GWB_CSS_BR_different_mu}, we show the GW power spectrum for different tensions for Case C with $p=10^{-1}$, which is the interesting case with a little enhancement of the GW power. The spectra are shown with sensitivity curves of various future experiments; SKA is the future pulsar timing array project, LISA and DECIGO are the future space-borne GW detectors, and Adv-LIGO describes the design sensitivity of the cross-correlation between four ground-based GW detectors (Advanced-LIGO, Advanced-VIRGO, and KAGRA).

\section{Summary}
\label{sec:summary}
In this paper, we studied how the distribution function of kinks on infinite cosmic superstrings are affected by a small reconnection probability and Y-junctions, and computed the power spectrum of the GW background from propagating kinks and kink-kink collisions. First, we calculated the correlation length and the velocity of cosmic superstrings using the extended VOS model, which enabled us to incorporate the formation of Y-junctions and the change of loop production efficiency due to a small reconnection probability.
Next, we added new terms in the time evolution equation of the kink distribution in order to take into account the fact that a kink entering a Y-junction generates three daughter kinks, who have smaller sharpness than the original one. 
We numerically solved the evolution equation and showed that the effect of Y-junction indeed reduces the sharpness of kinks and flattens the kink distribution function. 
Using the kink distribution, we calculated the GW background from propagating kinks and kink-kink collisions and found that kink-kink collisions tend to generate larger GW amplitudes.

In the case of $\mu_1:\mu_2:\mu_3=1:1:1$, we found that the amplitude of the GW background is always suppressed by Y-junctions because the amplitude of individual GW events depends on the kink sharpness and Y-junctions rapidly smooth out the kink sharpness. We also found that the GW amplitude becomes smaller for a smaller reconnection probability, since the number of Y-junctions increases when the reconnection probability is small and the effect of smoothing kink sharpness is enhanced. Although a small reconnection probability increases the number of strings inside the horizon and enhances the kink generation, we found that the effect of Y-junctions dominates and the GW amplitude is always reduced.
On the other hand, in the case of $\mu_1:\mu_2:\mu_3=1:10:10$, one of the daughter kinks inherits the original sharpness and the slope of the kink distribution is not flattened compared to the equal-tension case. We found that, when $p=10^{-1}$, the increase of the number of kinks by a small reconnection probability dominates the smoothing out of kink sharpness by Y-junctions, and the GW amplitude is slightly enhanced.

Naively, the GW background spectrum was expected to be larger in the case of cosmic superstrings, since the density of infinite strings becomes higher when the reconnection probability is small due to the low loop production efficiency. However, as summarized above, the GW amplitude from infinite cosmic superstrings turns out to be smaller than the one from ordinary infinite cosmic strings in the case of equal tensions, when we take into account the effect of Y-junctions. Our result suggests to reconsider our naive expectations that constraints on the tension of cosmic superstrings are tighter than the one for ordinary cosmic strings, and indicates that theories predicting cosmic superstrings with a large tension could still survive. 
This may also happen for the GW background from kinks on loops, which are considered to be larger than the GW background from infinite strings at high frequencies. Previous works \cite{Ringeval:2017eww,Jenkins:2018nty} predicted a large amplitude of the GW background from kink-kink collisions on loops. In particular \cite{Binetruy:2010cc} showed that the amplitude could be enhanced by Y-junctions since they increase the number of kinks on loops. However, the estimation was made by considering only sharp kinks $\psi\sim 1$, and the blunting of kinks was not taken into account. Consideration of the kink distribution, as performed in this work, would be necessary for a more accurate estimation of the GW background from loops with Y-junctions and it may give a smaller amplitude than expected, as happened in the case of infinite strings. We leave it for future work.

Note that the current upper bound on the GW background from pulsar timing array gives the strongest constraint on the cosmic string tension by considering GWs from loops.  Although GWs from infinite strings usually give a weaker constraint on string tension, it provides an independent bound and does not have any ambiguity on the initial loop size, which has been under debate and could weaken the constraint obtained by GWs from loops. 
  
Finally, we found that the shape of the power spectra of cosmic strings and cosmic superstrings are quite different. Thus, the spectral shape may be useful to distinguish between a cosmic-string or cosmic-superstring origin. This could be possible by GW searches in a wide range of frequencies such as CMB B-mode, pulsar timing arrays, as well as space-borne and ground-based direct detection experiments.

To demonstrate the characteristic effects of superstrings, we focused on the cases where the string tensions have a ratio of $1:1:1$ and $1:10:10$. Extending this work to explore the large parameter space of cosmic superstrings, one may be able to find cases where the GW amplitude and, consequently, the prospect of detection in future experiments are enhanced even more.

\acknowledgments
The work of KH was supported by a Grant-in-Aid for JSPS Research under Grant No.15J05029 and the Grant-in-Aid for Scientific Research No. 17K14282. The work of DN was supported by MEXT Grant-in-Aid for Scientific Research on Innovative Areas,~No.15H05890. SK is partially supported by JSPS KAKENHI No.17K14282 and Career Development Project for Researchers of Allied Universities.

\end{document}